\pdfoutput=1

\documentclass{vldb}

\newcommand{\subparagraph}{}

\usepackage[T1]{fontenc}
\usepackage{etoolbox,graphicx,setspace,listings,multicol,xspace,enumitem,booktabs,xcolor}
\usepackage{balance}
\usepackage{subcaption}
\usepackage{algorithm,mathtools}
\usepackage[noend]{algpseudocode}
\usepackage[skip=10pt,labelfont=bf]{caption}
\usepackage{tabularx}
\usepackage{times}
\allowdisplaybreaks

\usepackage{siunitx}
\newtoggle{arxiv}
\togglefalse{arxiv}

\iftoggle{arxiv}{
    \settopmatter{printacmref=false, printccs=false, printfolios=true}
    \renewcommand\footnotetextcopyrightpermission[1]{} 
    \pagestyle{plain} 
    \setcopyright{none} 
    \usepackage{mathtools}
    \usepackage{amsmath,amsthm}
}{
	\vldbTitle{Approximate Selection with Guarantees using Proxies}
	\vldbAuthors{Daniel Kang, Edward Gan, Peter Bailis, Tatsunori Hashimoto, Matei Zaharia}
	\vldbDOI{https://doi.org/10.14778/3407790.3407804}
	\vldbVolume{13}
	\vldbNumber{11}
	\vldbYear{2020}
	\usepackage[hyphens]{url}
	\usepackage{cite}
}

\newtheorem{theorem}{Theorem}
\newtheorem{lemma}{Lemma}

\iftoggle{arxiv}{
	\newtheorem*{defn*}{Definition}
	\newtheorem*{lemma*}{Lemma}
	\newtheorem{defn}{Definition}
}{
	\newdef{defn}{Definition}
}

\newcommand{\minihead}[1]{{\vspace{.45em}\noindent\textbf{#1.} }}

\newcommand{\specialcell}[2][c]{%
    \begin{tabular}[#1]{@{}c@{}}#2\end{tabular}}

\newcommand*{\sn}{\textsc{SUPG}\xspace}
\newcommand*{\noscope}{\textsc{NoScope}\xspace}

\newcommand*{\ci}{\textsc{CI}\xspace}
\newcommand*{\noci}{\textsc{NoCI}\xspace}
\newcommand*{\uniform}{\textsc{U}\xspace}
\newcommand*{\imp}{\textsc{IS}\xspace}

\newcommand*{\rt}{\textsc{R}\xspace}
\newcommand*{\pt}{\textsc{P}\xspace}

\newcommand*{\AlgAP}{\uniform-\noci-\pt}
\newcommand*{\AlgAR}{\uniform-\noci-\rt}
\newcommand*{\AlgBP}{\uniform-\ci-\pt}
\newcommand*{\AlgBR}{\uniform-\ci-\rt}
\newcommand*{\AlgCP}{\imp-\ci-\pt}
\newcommand*{\AlgCR}{\imp-\ci-\rt}

\newcommand*{\universe}{\mathcal{D}}
\newcommand*{\SD}{\mathcal{D}}
\newcommand*{\SQ}{\mathcal{Q}}
\newcommand*{\SSe}{\mathcal{S}}
\newcommand*{\tth}{\tau}
\newcommand*{\res}{\mathcal{R}}
\newcommand*{\tm}{\gamma}

\newcommand*{\ex}{\mathop{\mathbb{E}}}
\newcommand*{\var}{\mathop{\text{Var}}}

\newtoggle{rcolors} \togglefalse{rcolors}
\newcommand{\colora}[1]{\iftoggle{rcolors}{{\color{red}{#1}}}{#1}}

\lstdefinelanguage{SQL}{
 keywords={AND, OR, WHERE, SELECT, FROM, LIKE, GROUP, BY, NOT, COUNT, DISTINCT, DIFF, BETWEEN},
 keywordstyle=\color{black},
 ndkeywords={NULL, true, false},
 ndkeywordstyle=\bfseries,
 basicstyle=\small\ttfamily,
 identifierstyle=\color{black},
 sensitive=false,
 comment=[l]{--},
 morecomment=[s]{/*}{*/},
 commentstyle=\color{NavyBlue}\ttfamily,
 string=[s]{"}{"},
 morestring=[s]{`}{'},
 showstringspaces=false,
 stringstyle=\color{violet}\ttfamily,
}
\AtBeginDocument{%
  \providecommand\BibTeX{{%
    \normalfont B\kern-0.5em{\scshape i\kern-0.25em b}\kern-0.8em\TeX}}}

\begin{document}

\iftoggle{arxiv}{
	\title{Approximate Selection with Guarantees using Proxies}
	\titlenote{Marked authors contributed equally. Preprint under review.}
	\author{Daniel Kang*, Edward Gan*, Peter Bailis, Tatsunori Hashimoto, Matei Zaharia}
	\affiliation{
	    \institution{Stanford University}
	}
	\begin{abstract}

Due to the falling costs of data acquisition and storage, researchers
and industry analysts often want to find all instances of rare
events in large datasets.
For instance, scientists can cheaply capture thousands of hours of video, but
are limited by the need to manually inspect long videos to identify
relevant objects and events.
To reduce this cost, recent work proposes to use cheap proxy models, such as
image classifiers, to identify an approximate set of data points satisfying a
data selection filter.
Unfortunately, this recent work does not provide the statistical accuracy
guarantees necessary in scientific and production settings.

In this work, we introduce novel algorithms for approximate selection queries
with \emph{statistical accuracy guarantees}.
Namely, given a limited number of exact identifications from an oracle, often a
human or an expensive machine learning model, our algorithms meet a minimum
precision or recall target with high probability.
In contrast, existing approaches can catastrophically fail in satisfying these
recall and precision targets.
We show that our algorithms can improve query result quality by up to 30$\times$
for both the precision and recall targets in both real and synthetic datasets.

\end{abstract}

	\maketitle
}{
	\title{Approximate Selection with Guarantees using Proxies}
	\numberofauthors{1}
	\author{
	  Daniel Kang\thanks{Authors contributed equally}, Edward Gan\footnotemark[1],
    Peter Bailis, Tatsunori Hashimoto, Matei Zaharia \\
	  \affaddr{Stanford University}\\
    \texttt{supg@cs.stanford.edu}
	}
	\maketitle
	\begin{abstract}

Due to the falling costs of data acquisition and storage, researchers
and industry analysts often want to find all instances of rare
events in large datasets.
For instance, scientists can cheaply capture thousands of hours of video, but
are limited by the need to manually inspect long videos to identify
relevant objects and events.
To reduce this cost, recent work proposes to use cheap proxy models, such as
image classifiers, to identify an approximate set of data points satisfying a
data selection filter.
Unfortunately, this recent work does not provide the statistical accuracy
guarantees necessary in scientific and production settings.

In this work, we introduce novel algorithms for approximate selection queries
with \emph{statistical accuracy guarantees}.
Namely, given a limited number of exact identifications from an oracle, often a
human or an expensive machine learning model, our algorithms meet a minimum
precision or recall target with high probability.
In contrast, existing approaches can catastrophically fail in satisfying these
recall and precision targets.
We show that our algorithms can improve query result quality by up to 30$\times$
for both the precision and recall targets in both real and synthetic datasets.

\end{abstract}

}

\section{Introduction}
\label{sec:intro}

As organizations are now able to collect large datasets, they regularly aim to
find all instances of rare events in these datasets.
For example, biologists in a lab at Stanford have
collected months of video of a flower field and wish to identify timestamps when
hummingbirds are feeding so they can match hummingbird feeding patterns with
microbial readings from the flowers. Furthermore, our contacts at an autonomous
vehicle company are interested in auditing when their labeled data may be wrong,
e.g., missing pedestrians~\cite{dywer2020popular}, so they can correct them.
Other recent work has also studied this problem~\cite{kang2017noscope,
lu2018accelerating, anderson2018predicate, hsieh2018focus}.
Importantly, these events are \emph{rare} (e.g., at most 0.1-1\% of frames
contain hummingbirds) and users are interested in the \emph{set of matching
records} as opposed to aggregate measures (e.g., counts).

Unfortunately, executing \emph{oracle predicates} (e.g., human labelers or deep
neural networks) to find such events can be prohibitively expensive, so many
applications have a \emph{budget} on executing oracle predicates. For example,
biologists can watch only so many hours of video and companies have fixed
labeling budgets.

To reduce the cost of such queries, recent work, such as \noscope and
probabilistic predicates~\cite{kang2017noscope, lu2018accelerating,
anderson2018predicate, hsieh2018focus}, has proposed to use cheap \emph{proxy}
models that approximate ground truth \emph{oracle} labels (e.g., labels from
biologists).  These proxy models are typically small machine learning models
that provide a confidence score for the label and selection predicate. If the
proxy model's confidence scores are reliable and consistent with the oracle,
they can be used to filter out the vast majority of data unlikely to match.

There are two major challenges in using these proxy models to reduce the
labeling cost subject to a budget: reliability of proxy models and oracle labeling efficiency.

\begin{figure}
  \includegraphics[width=\columnwidth]{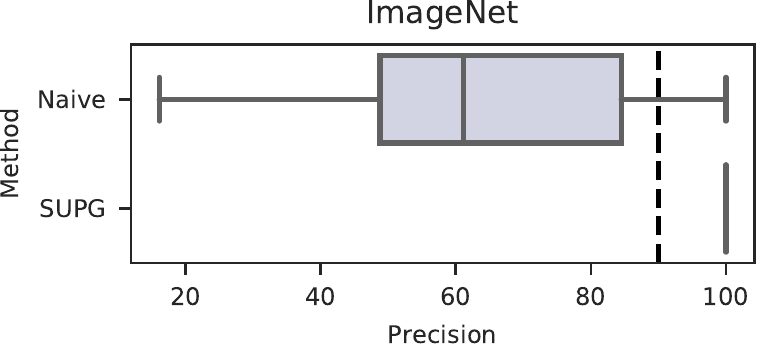}
  \caption{Box plot of achieved precisions of naive sampling
  from recent work \cite{kang2017noscope, lu2018accelerating} and our improved
  algorithm. Over 100 runs targeting a precision of 90\%, the naive algorithm
  returns precisions as low as 65\% for over half the runs. In contrast, our
  algorithms (SUPG) achieve the precision target with high probability.}
  \label{fig:intro-guarantees}
  \vspace{-1em}
\end{figure}

First, given the budget, using an unreliable proxy model can result in false negatives
or positives, making it difficult to guarantee the accuracy of query results.
Existing systems do not provide guarantees on the accuracy. In fact they can
fail unpredictably and catastrophically, providing results with low accuracy a
significant fraction of the time~\cite{kang2017noscope, kang2019blazeit,
lu2018accelerating, canel2019scaling, anderson2018predicate, hsieh2018focus}.
For example, when users request a precision of at least 90\%, over repeated
runs, existing systems return results with less than 65\% precision \emph{over
half the time}, with some runs low as 20\% (Figure~\ref{fig:intro-guarantees}).
These failures can be even worse in the face of shifting data distributions,
i.e., model drift (Section~\ref{sec:eval-guarantees}). Such failures are
unacceptable in production deployment and for scientific inference.

Second, existing systems do not make efficient use of limited oracle labels to
maximize the quality of query results. To avoid vacuous results (e.g., achieving
a perfect recall by returning the whole dataset will have poor precision),
\noscope, probabilistic predicates, and other recent work uniformly sample
records to label with the oracle in order to decide on the final set of records
to return. We show that this is wasteful. In the common case where records
matching the predicate are rare, the vast majority of uniformly sampled records
will be negatives, with too few positives. Thus, naively extending existing
techniques to have accuracy guarantees can fail to maintain high result quality
given these uninformative labels.

In response we develop novel algorithms that provide \emph{both} statistical
guarantees and efficient use of oracle labels for approximate selection. We
further develop query semantics for the two settings we consider: the recall
target and precision target settings.

\minihead{Accuracy guarantees}
To address the challenge of guarantees on failure probability, we first define
probabilistic guarantees for two classes of approximate selection queries. We
have found that users are interested in queries targeting a minimum recall (RT
queries) or targeting a minimum precision (PT queries), subject to an oracle
label budget and a failure probability on the query returning an invalid result
(Section~\ref{sec:queries}). For instances, the biologist are interested in 90\%
recall and a failure probability of at most 5\%.

We develop novel algorithms (\sn algorithms) that
provide these guarantees by using the oracle budget to randomly sample records
to label, and estimating a proxy confidence threshold $\tth$ for which it
is safe to return all records with proxy score above $\tth$. Naive use of
uniform sampling will not account for the probability that there is a deviation
between observed labels and proxy scores, and will further introduce multiple
hypothesis testing issues. This will result in a high probability
of failure. In response, we make careful use of confidence intervals and multiple
hypotheses corrections to ensure that the failure probability is controlled.

\minihead{Oracle sample efficiency}
A key challenge is deciding which data points to label with the oracle given the
limited budget: as we show, uniform sampling is inefficient. Instead, we develop
novel, optimal importance sampling estimators that use of the correlation
between the proxy and the oracle, while taking into account possible
mismatches between the binary oracle and continuous proxy. Intuitively, importance
sampling upweights the probability of sampling data points with high proxy
scores, which are more likely to contain the events of interest.

However, naive use of importance sampling results in poor performance when
sampling according to proxy scores. Using a variance decomposition, we find
that a standard approach for obtaining importance weights (i.e., using weights
proportional to the proxy) is suboptimal and, excluding edge cases, performs no
better than uniform random sampling.

Instead, we show that sampling proportional to the \emph{square root} of the proxy 
scores allows for more efficient estimates of the proxy threshold
when the proxy scores are confident and reliable (Section~\ref{sec:stat_efficiency}). 
For precision target queries, we additionally extend importance
sampling to use a two-stage sampling procedure. In
the first stage, our algorithm estimates a safe interval to further sample. In
the second stage, our algorithm samples directly from this range, which we show
greatly improves sample efficiency.

Careless of use of importance sampling can hurt result quality when used with
poor proxy models. If proxy scores are uncorrelated with the true labels,
importance sampling will in fact increase the variance of sampling. To address
these issues, we defensively incorporate uniform samples to guard against
situations where the proxy may be adversarial \cite{owen2000safeis}. This
procedure still maintains the probabilistic accuracy guarantees.

\vspace{0.5em}

We implement and evaluate these algorithms on real and synthetic datasets and
show that our algorithms achieve desired accuracy guarantees, 
even in the presence of model drift. We further show
that our algorithms outperform alternative methods in providing higher result quality,
by as much as 30$\times$ higher recall/precision under precision/recall
constraints respectively.

In summary, our contributions are:
\begin{enumerate}[topsep=0em,itemsep=0em,parsep=0em]
  \item We introduce semantics for approximate selection queries with
  probabilistic accuracy guarantees under a limited oracle budget.

  \item We develop algorithms for approximate selection queries that satisfy
  bounded failure probabilities while making efficient use of the oracle labels.

  \item We implement and evaluate our algorithms on six real-world and
  synthetic datasets, showing up to 30$\times$ improvements in quality and
  consistent ability to achieve target accuracies.

\end{enumerate}

\section{Use Cases}
\label{sec:use-case}
To provide additional context and motivation for approximate selection queries, we describe scenarios where
statistically efficient queries with guarantees are essential. 
Each scenario is informed by discussions with academic and industry collaborations.

\subsection{Biological Discovery}
\begin{figure}
\centering
  \begin{subfigure}[h]{0.45\linewidth}
    \includegraphics[width=0.99\columnwidth]{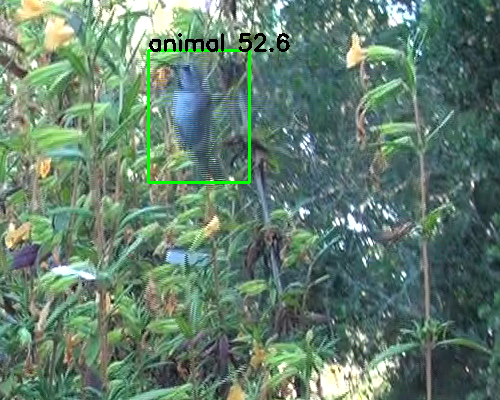}
    \caption{Hummingbird present}
  \end{subfigure}
  \quad
  \begin{subfigure}[h]{0.45\linewidth}
    \includegraphics[width=0.99\columnwidth]{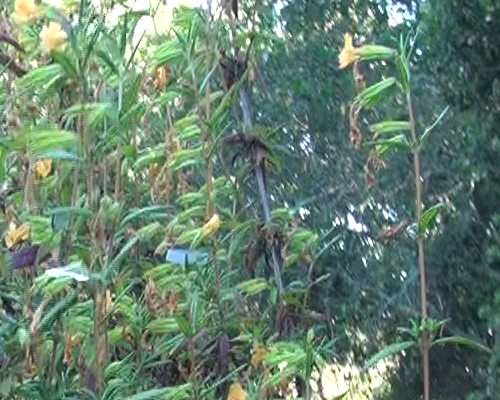}
    \caption{No hummingbirds}
  \end{subfigure}
  \caption{Sample matching (a) and non-matching (b) frames for a selection query over a video stream used by our biologist collaborators. 
  Only a small ($<.1\%$) fraction of frames have hummingbirds present and match the filter predicate, 
  making manual inspection difficult. 
  DNNs can serve as proxies to identify hummingbirds as shown in
  (a), but the confidence scores can be unreliable.}
  \label{fig:hummingbirds}
\end{figure}

\minihead{Scenario}
We are actively collaborating with the Fukami lab at Stanford University,
who study bacterial colonies in flowers~\cite{san2020land}. 
The Fukami lab is interested in hummingbirds that move bacteria between flowers as they feed, as this bacterial movement
can affect both the micro-ecology of the flowers and later
hummingbird feeding patterns.
To study such phenomena, they have collected videos of bushes with
tagged flowers at the Jasper Ridge biological preserve. 
They have recorded 
six views of the scene with a total of approximately 9 months of video. 
At 60 fps, this is approximately 1.4B frames of video. 
To perform downstream analyses, our collaborators want to select all 
frames in the video that contain hummingbirds.
Due to the rarity of hummingbird appearances ($<0.1\%$) and the length of the video,
they have been unable to manually review the video in its entirety.
We illustrate the challenge in Figure~\ref{fig:hummingbirds}.

\minihead{Proxy model}
Prior to our collaboration, the Fukami lab used motion detectors as a proxy
for identifying frames with birds. 
However, the motion detectors have severe limitations: their
precision is extremely low (approximately 2\%) and they do not cover the full field of view of the bush.
As an alternative, we are actively using DNN object detector models to identify
hummingbirds directly from frames of the video \cite{he2017mask,
beery2019efficient}.
These DNN models are more precise than motion detectors, and can provide
a confidence score in addition to a Boolean predicate result.

During discussion with the Fukami lab, we have found that the scientists require
high probability guarantees on recall, as finding the majority of
hummingbirds is critical for downstream analysis. Furthermore, they are
interested in improving precision relative to the motion detectors. The
scientists have specified that they need a recall of at least 90\% and a
precision that is as high as possible, ideally above 20\%.

\begin{figure}[t!]
\begin{lstlisting}[frame=single, xrightmargin=0.8em, xleftmargin=0.7em]
SELECT * FROM table_name
  WHERE filter_predicate
  ORACLE LIMIT o
  USING proxy_estimates
  [RECALL | PRECISION] TARGET t 
  WITH PROBABILITY p
\end{lstlisting}
\vspace{-1em}
  \caption{Syntax for specifying approximate selection queries. Users provide
  a precision or recall target, a budget for the number of oracle predicate evaluations, and a probability of success.
  \label{fig:syntax}}
  \vspace{-0.5em}
\end{figure}

\subsection{Autonomous Vehicle Training}
\minihead{Scenario}
An autonomous vehicle company may collect data in a new area. To train the DNNs
used in the vehicle, the company may extract point cloud or visual data and
use a labeling service to label pedestrians. Unfortunately, labeling services
are known to be noisy and may not label pedestrians even when they are
visible~\cite{dywer2020popular}.

To ensure that all pedestrians are labeled, an analyst may wish to select all frames
where pedestrians are present but are not annotated in the labeled data.
However, as autonomous vehicle fleets collect enormous amounts of data (petabytes per day), the analyst is not able to manually inspect all the data.

\minihead{Proxy model}
As the proxy model, the analyst can use an object detection method and remove
boxes that are in the labeled dataset. The analyst can then use the confidences
from the remaining boxes from the object detection as the proxy scores.

As this is a mission-critical setting, the analyst is interested in guarantees
on recall. Missing pedestrians in the labeled dataset can transfer to missing
pedestrians at deployment time, which can cause fatal accidents.

We further note that the analyst may also be interested in using other proxies,
such as 3D detections from LIDAR data. In this work, we only study the use of a
single proxy model, but we see extending our algorithms to multiple proxy models
as an exciting area of future work (Section~\ref{sec:discussion}).

\vspace{0.75em}

We note that this scenario is not limited to autonomous vehicles but can apply
to other scenarios where curating high quality machine learning datasets is of
paramount concern.

\subsection{Legal Discovery and Analysis}
\minihead{Scenario}
Lawyers are often tasked with analyzing large corpora of data, e.g.,
they may inspect a corpus of emails as part of legal discovery, or may
be tasked to analyze if private information was leaked in a data breach~\cite{exterro, textiq}.
This process is expensive as hiring contract lawyers to manually inspect
documents is time consuming and expensive. As a result, a number of companies
are interested in leveraging automatic methods to select all records that
match sensitive named entities or reference relevant legal concepts.

\minihead{Proxy model}
As the proxy model, an analyst may fine-tune a sophisticated language
understanding model, such as BERT~\cite{devlin2018bert}. The analyst can deploy
this model over the corpus of text data and extract noisy labels to help the
lawyers. For sensitive issues, companies would benefit from tools that can
provide either recall or precision guarantees depending on the scenario.

%
%

\section{Approximate Selection Queries}
\label{sec:queries}

\sloppypar{
We introduce our definitions for our approximate selection queries (\sn
queries), describe the probabilistic
guarantees they respect, and define metrics for comparing the quality
of the results.
}

\subsection{Query Semantics}
\label{sec:query-semantics}

A \sn query is a selection query for set of records matching a predicate, with syntax given in Figure~\ref{fig:syntax}.
Unlike much of the existing work in approximate query processing,
\sn queries return a set of matching records rather than a scalar or vector aggregate
\cite{hellerstein1997online, agarwal2013blinkdb}. We defined these semantics to
formalize a common class of queries our collaborators and industrial contacts are
interested in executing.

The query specifies a filter predicate given by a ``ground truth'' oracle, as
well as a limited budget of total calls to the oracle over the course of query
execution.
\colora{
We use the term oracle to refer to any expensive predicate the user wishes to
approximate. In some cases, the oracle may be an expensive DNN (e.g., the highly
accurate Mask R-CNN~\cite{he2017mask}) that may not exactly match the ground
truth labels that a human labeler would provide. However, the use of proxies to
approximate powerful deep learning models is common in the
literature~\cite{kang2017noscope, kang2019blazeit, lu2018accelerating,
hsieh2018focus, anderson2018predicate}, so we study how to provide guarantees in
applications that use a larger DNN as an oracle.
}

Since oracle usage is limited, queries also specify proxy confidence
scores for whether a record matches the predicate. The proxy scores must be
correlated with the probability that a record matches the filter predicate to be
useful. Nonetheless, our novel algorithms will
return valid results even if proxy scores are not correlated.

The accuracy of the set of results can be measured using either
\emph{recall} (the fraction of true matches returned) or \emph{precision} (the fraction
of returned results that are true matches).
Based on the application, a user can specify either a minimum recall or precision target
as well as a desired probability of achieving this target.
We refer to these two options as \emph{precision target} (PT) 
and \emph{recall target} (RT) queries.
As an example, consider the following RT query:
\begin{lstlisting}[frame=none]
SELECT * FROM hummingbird_video
  WHERE HUMMINGBIRD_PRESENT(frame) = True
  ORACLE LIMIT 10,000
  USING DNN_CLASSIFIER(frame) = "hummingbird"
  RECALL TARGET 95%
  WITH PROBABILITY 95%
\end{lstlisting}
where both \texttt{HUMMINGBIRD\_PRESENT} and \texttt{DNN\_CLASSIFIER} are
user-defined functions (UDFs).
This query selects the frames of the video that contains a hummingbird with
recall at least 95\%, using at most 10,000 oracle evaluations,
and a failure probability of at most 5\%, using confidence probabilities from a 
DNN classifier as a proxy. The oracle could be a human labeler or expensive DNN.

Finally, we note that some queries may require both a recall and precision target.
Unfortunately, jointly achieving both targets may require an unbounded number of
oracle queries. Since all use cases we consider have limited budgets, we defer
our discussion of these queries to
\iftoggle{arxiv}{
Appendix~\ref{sec:jt}.
}{
an extended version of this paper~\cite{kang2020approximate}.
}

\subsection{Probabilistic Guarantees}
More formally, a \sn query $\SQ$ is defined by an oracle predicate 
$O(x) \in \{0,1\}$ over a set of records $x$ from a dataset $\universe$.
The ideal result for the query would be the matching records
$O^{+} \coloneqq \{x \in \universe : O(x)=1\}$.
However, since the oracle is assumed to be expensive, 
the query specifies a budget of $s$ calls to the oracle $O(x)$, 
as well as a proxy model $A(x) \in [0,1]$ whose use is unrestricted.
The query specifies a minimum recall or precision target $\tm$.
Then, a \emph{valid} query result would be a set of records $\res$ such that
$\text{Precision}(\res) > \tm_{p}$ or $\text{Recall}(\res) > \tm_{r}$ depending
on the query type.
Recall that 
\begin{align*}
\text{Precision}(\res) \coloneqq \frac{|\res \cap O^{+}|}{|\res|} \qquad
\text{Recall}(\res) \coloneqq \frac{|\res \cap O^{+}|}{|O^{+}|}.
\end{align*}

A \sn query further specifies a failure probability $\delta$. 
Many precision or recall targets $\tm$ may be impossible to achieve deterministically
given a limited budget of $s$ calls to the oracle, as they require exhaustive
search.
Thus, it is common in approximate query processing and statistical inference to use
randomized procedures with a bounded failure probability \cite{cormen2009introduction}.
A randomized algorithm satisfies the guarantees in $\SQ$ if it produces valid results $\res$
with high probability. 
That is, for PT queries:
\begin{equation}
\Pr[\text{Precision}(\res) \geq \tm_{p}] \geq 1-\delta
\end{equation}
and for RT queries:
\begin{equation}
\Pr[\text{Recall}(\res) \geq \tm_{r}] \geq 1-\delta.
\label{eqn:recall_guarantee}
\end{equation}

These high probability guarantees are much stronger than merely achieving 
an average recall or precision as many existing systems do
\cite{kang2017noscope, lu2018accelerating, anderson2018predicate, hsieh2018focus}.
For example, in Figure~\ref{fig:naive-recall} we illustrate the true recall provided for queries
targeting 90\% recall to \noscope system, and compare them with the recall provided
by \sn which satisfies the stronger guarantee in Equation~\ref{eqn:recall_guarantee}.
\noscope only achieves the target recall approximately half of the time, with many runs
failing to achieve the recall target by a significant margin.
Such results that fail to achieve the recall target would have a significant
negative impact on downstream statistical analyses.

\subsection{Result Quality}
Since \sn queries only specify a target for either precision or recall (the \emph{target metric}),
there are many valid results for a given query which may be more or less useful.
For instance, if a user targets $99\%$ recall the entire dataset is always a valid
result, even though this may not be useful to the user.
In this case, it would be more useful to return a smaller set of records to minimize false positives.
Similarly, if a user targets high precision the empty set is always a valid result,
and is equally useless.
Thus, we define selection query quality in this paper as follows:
\begin{defn}
For RT/PT queries, a higher quality result is one with higher precision/recall,
respectively.
\end{defn}
There is an inherent trade-off between returning valid results and maximizing result quality,
analogous to the trade-off between maximizing precision and maximizing recall in 
binary classification \cite{gordon1989recall, buckland1994relationship},
but efficient use of oracle labels will allow us to develop more efficient importance sampling based query techniques.


\section{Algorithm Overview}
\label{sec:sys_arch}
In this section, we describe the system setting that our \sn algorithms
operate in, and outline the major stages in the algorithm:
sampling oracle labels, choosing a proxy threshold, and returning
a set of data record results.

\subsection{Operational Architecture}
Our algorithms are designed for batch query systems
that perform selection on datasets of existing records. 
Users can issue queries over the data with specified predicates and parameters
as described earlier.
Note that the oracle and proxy models used to evaluate the filter predicate 
are provided by the user as UDFs (callback functions) 
and are not inferred by the system.
Thus, a user must provide either a ground truth DNN or interface to obtain human input as an
oracle, as well as pre-trained inexpensive proxy models.
In practice, one can provide user interfaces for interactively requesting human
labels~\cite{scale}
as well as scripts for automatically constructing smaller proxy models from an
existing oracle~\cite{kang2017noscope, lu2018accelerating},
though those are outside the scope of this paper.

\begin{figure}
  \includegraphics[width=\columnwidth]{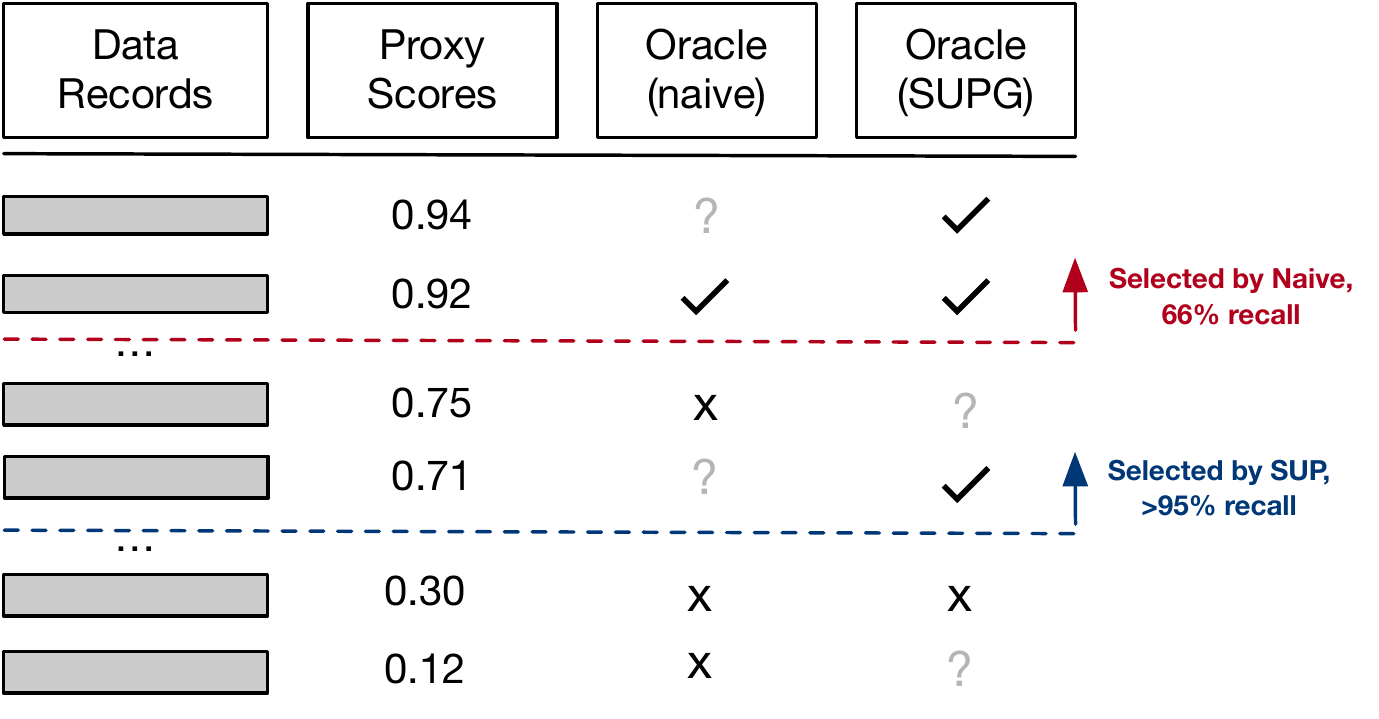}
  \caption{Our \sn algorithms uses sampled oracle labels and proxy scores to identify a subset of records
  that satisfy a recall or precision target with high probability.
  Naive selection methods, as used by recent work, would make less efficient use of limited oracle labels, and provide
  weaker guarantees: in fact they often fail to achieve a target recall or precision.}
  \label{fig:sys-arch}
\end{figure}

We illustrate how \sn uses oracle and proxy models in Figure~\ref{fig:sys-arch}.
For all query types, \sn first executes the proxy model over the complete set of records $\SD$
as we assume the proxy model is cheap relative to the oracle model. 
Then, \sn samples a set $\SSe$ of $s$ records to label using the oracle model.
The choice of which records to label using the oracle is done adaptively, that is, the
choice of samples may depend on the results of previous oracle calls for a given query.

\begin{algorithm}
\begin{algorithmic}
\Function{\sn{Query}}{$\SD$, $A$, $O$}
  \State $\SSe \gets \texttt{SampleOracle}(\SD)$
  \State $\tth \gets \texttt{EstimateTau}(\SSe)$
  \State $\res_1 \gets \{x : x \in \SSe \wedge O(x)=1\}$
  \State $\res_2 \gets \{x : x\ \in \SD \wedge A(x) \geq \tth\}$
  \State \textbf{return} $\res_1 \cup \res_2$
\EndFunction
\end{algorithmic}
\caption{\sn query processing}
\label{alg:query-process}
\end{algorithm}
We summarize this sequence of operations \sn uses to return query results
in Algorithm~\ref{alg:query-process}.
After calling the oracle to obtain predicate labels over a sample $\SSe$,
\sn sets a proxy score threshold $\tth$
and then returns results $\res$ that consist of both
labeled records in $\SSe$ matching the oracle predicate
as well as records with proxy scores above the threshold $\tth$.
$\tth$ is tuned so that
the final results $\res$ satisfy minimum recall or precision targets with high probability, 
and we describe the process for setting $\tth$ below.

\subsection{Choosing a Proxy Threshold}
Since the proxy scores are the only source of information on the query predicate besides
the oracle model, \sn naturally returns records corresponding to all
records with scores above a threshold $\tth$. This strategy is known to be optimal
in the context of retrieval and ranking as long as proxy scores grow monotonically
with an underlying probability that the record matches a predicate~\cite{manning2008IR}.
We have observed in practice that this is approximately true for proxy models
by computing empirical match rates for bucketed ranges of the proxy scores, so we
use this as the default strategy in \sn. 
For proxy models that are completely uncorrelated or have non-monotonic
relationships with the oracle, all algorithms using proxies will have
increasingly poor quality, but \sn will still provide accuracy guarantees.

Thus, the key task is selecting the threshold $\tth$ to maintain
result validity while maximizing result quality.
This threshold must be set at query time since the relation between proxy scores
and the predicate is unknown, especially when production model drift is an issue.
Existing systems have often relied on pre-set thresholds determined ahead of time,
which we show in Section~\ref{sec:eval-guarantees} can lead to severe violations of result validity.

One naive strategy for selecting $\tth$ at query time is to uniform randomly sample records
to label with the oracle until the budget is exhausted, and then select $\tth$ that
achieves a target accuracy over the sample.
However, this strategy on its own does not provide strong accuracy guarantees or make efficient use of the
sample budget.
Thus, in Section~\ref{sec:alg} we introduce more sophisticated methods for sampling
records and estimating the threshold: that is, implementations of
$\texttt{SampleOracle}$ and
$\texttt{EstimateTau}$.

\begin{table}
\centering
\caption{Notation Summary}
\label{table:notation}
\vspace{-0.5em}
\begin{tabular}{ll}
Symbol & Description \\\hline
$O(x)$ & Oracle predicate value\\
$A(x)$ & Proxy confidence score \\
$\delta$ & Failure probability \\
$\tm$ & Target Recall / Precision\\
$\tth$ & Proxy score threshold for selection \\
$\SSe$ & Records sampled for oracle evaluation \\
\end{tabular}
\vspace{-0.5em}
\end{table}

\section{Estimating proxy thresholds}
\label{sec:alg}

Recall that \sn selects all records with proxy scores above a threshold
$\tth$.
Denote this set of records 
$$\SD(\tth) \coloneqq \{x : A(x) \geq \tth\}.$$
\sn query accuracy thus critically depends on the choice of $\tth$.
In this section we describe our algorithms for estimating a threshold
that can guarantee valid results with high probability, while maximizing
result quality. 
While precision target (PT) and recall target (RT) queries require slightly
different threshold estimation routines, in both cases \sn samples records to 
label with the oracle. Using this sample, \sn will
select a threshold $\tth$ that achieves the target metric on the dataset $\SD$
with high probability.

In order to explain our algorithms and compare them with existing work,
we will also describe a number of baseline techniques which do not provide
statistical guarantees, and do not make efficient use of oracle labels
to improve result quality.

We now describe baselines without guarantees, how
to correct these baselines for statistical guarantees on failure probability,
and finally our novel importance sampling algorithms.

\subsection{Baselines Without Guarantees}
The simplest strategy for estimating a valid threshold would be to
take a uniform i.i.d.~random sample of records $\SSe$, label the records
with the oracle, and then use $\SSe$ as an exact representative of the dataset
$\SD$ when choosing a threshold.
This is the approach used by probabilistic predicates and \noscope
\cite{kang2017noscope, lu2018accelerating},
and we call this approach \uniform-\noci because it uses a uniform
sample and does not account for failure probabilities using confidence intervals (CI).
Let $\text{Recall}_{\SSe}(\tth)$ and
$\text{Precision}_{\SSe}(\tth)$ denote the empirical
recall and precision for the sampled data $\SSe$, and $1_c$ denote the indicator function that condition $c$ holds:
 \begin{align}
\text{Recall}_{\SSe}(\tth) &\coloneqq \frac{\sum_{x \in \SSe} 1_{A(x) \geq \tth} O(x)}{\sum_{x \in \SSe} O(x)} \\
\text{Precision}_{\SSe}(\tth) &\coloneqq \frac{\sum_{x \in \SSe} 1_{A(x) \geq \tth} O(x)}{|\SSe|}.
\end{align}

The \AlgAP approach maximizes result quality subject to constraints on these empirical
recall and precision estimates. For PT queries this results in finding the minimal $\tau$ (minimizing false negatives) that achieves the target metric on $\SSe$, and
for RT queries this results in finding the maximum $\tau$ (minimizing false positives).
Formally this is defined as,
\begin{align}
\tth_{\text{\AlgAP}}(\SSe) &= \min \{\tth : \text{Precision}_{\SSe}(\tth) \geq \tm \} \\
\tth_{\text{\AlgAR}}(\SSe) &= \max \{\tth : \text{Recall}_{\SSe}(\tth) \geq \tm \}.
\end{align}
However, we have no guarantee that the thresholds selected in this way will
provide valid results on the complete dataset, due to the
random variance in choosing a threshold based on the limited sample.
We empirically show that such algorithms fail to achieve targets up to 80\% of
the time in Section~\ref{sec:eval-guarantees}.

\subsection{Guarantees through Confidence Intervals}
In order to provide probabilistic guarantees, we form confidence intervals over $\tth$
and take the appropriate upper or lower bound.

\minihead{Normal approximation}
In Lemma~\ref{lem:clt} we describe an asymptotic bound relating sample averages to
population averages, allowing us to bound the discrepancy between recall
and precision achieved on $\SSe$ vs $\SD$. This approximation is
commonly used in the approximate query processing
literature~\cite{hellerstein1997online, haas1999ripple, agarwal2014knowing}.

For ease of notation we will refer
to the upper and lower bounds provided by Lemma~\ref{lem:clt} using helper functions
\begin{align}
\text{UB}(\mu,\sigma,s,\delta) &\coloneqq \mu + \frac{\sigma}{\sqrt{s}}\sqrt{2\log{\frac{1}{\delta}}} \\
\text{LB}(\mu,\sigma,s,\delta) &\coloneqq \mu - \frac{\sigma}{\sqrt{s}}\sqrt{2\log{\frac{1}{\delta}}}.
\end{align}

\begin{lemma}
Let $\SSe$ be a set of $s$ i.i.d.~random variables $x \sim \mathcal{X}$ with mean $\mu$ and
finite variance $\sigma^2$ and sample mean $\hat{\mu}$.
Then,
\begin{align*}
\lim_{s\rightarrow \infty} \Pr\left[
\hat{\mu} \geq \text{UB}(\mu,\sigma,s,\delta)
\right] &\leq \delta
\end{align*}
and
\begin{align*}
\lim_{s\rightarrow \infty} \Pr\left[
\hat{\mu} \leq \text{LB}(\mu,\sigma,s,\delta)
\right] &\leq \delta.
\end{align*}
\label{lem:clt}
\end{lemma}

Lemma~\ref{lem:clt} defines the expected variation in recall and precision estimates
as $s$ grows large, and follows from the Central Limit Theorem \cite{wasserman2013all}.
Using this, we can select conservative thresholds that with
high probability still provide valid results on the underlying dataset $\SD$.
\colora{Though this bound is an asymptotic result for large $s$, quantitative convergence rates for such statistics
are known to be fast \cite{bentkus1996berry} and we found that this approach provides the appropriate
probabilistic guarantees at sample sizes $s > 100$.}

\minihead{Other confidence intervals}
\colora{
Throughout, we use Lemma~\ref{lem:clt} to compute confidence intervals. There
are other methods to compute confidence intervals, e.g., the
bootstrap~\cite{efron1992bootstrap}, Hoeffding's
inequality~\cite{hoeffding1994probability}, and ``exact'' methods for Binomial
proportions (Clopper-Pearson interval)~\cite{clopper1934use}. We show that the
normal approximation matches or outperforms alternative methods of computing
confidence intervals (Section~\ref{sec:eval-sensitivity}). Since the normal
approximation is straightforward to implement and applies to both uniform and
importance sampling we use it throughout.
}

\vspace{0.3em}
We will now describe baseline uniform sampling based methods for estimating $\tth$ in both RT and PT queries.

\subsubsection{Recall Target}
For recall target queries, we want to estimate a threshold $\tth$ such that 
$\text{Recall}_{\SD}(\tth) \geq \tm$
with probability at least $1-\delta$.
To maximize result quality we would further like to
make $\tth$ as large as possible.
We present the pseudocode for a threshold selection routine \AlgBR that provides
guarantees on recall in Algorithm~\ref{alg:thresh-rt}.

Note that Algorithm~\ref{alg:thresh-rt} finds a cutoff $\tth$ that achieves a conservative recall of
$\tm'$ on $\SSe$ instead of the target recall $\tm$. This inflated recall target accounts for the potential random variation from forming the threshold on $\SSe$ rather than $\SD$.

\minihead{Validity justification}
Let $\tth_o$ be the largest threshold providing valid recall on $\SD$:
$$\tth_o \coloneqq \max \{\tth : \text{Recall}_{\SD}(\tth) \geq \tm \}$$
If $\text{Recall}_{\SSe}(\tth_o) \leq \tm'$ then Algorithm~\ref{lem:clt} will 
select a threshold $\tth'$ where $\tth' \leq \tth_o$ 
since recall varies inversely with the threshold.
Then, $\text{Recall}_{\SD}(\tth') \geq \text{Recall}_{\SD}(\tth) \geq \tm$
and the results derived from $\tth'$ would be valid.

It remains to show that with probability $1-\delta$, $\tm'$ satisfies:
\begin{equation}
\text{Recall}_{\SSe}(\tth_o) \leq \tm'.
\label{eqn:imp_rt_toshow}
\end{equation}

Let $Z_1(\tth),Z_2(\tth)$ be sample indicator random variables for
matching records above and below $\tth_o$, corresponding to the
samples in $\SSe$.
\begin{align*}
Z_1(\tth) &\coloneqq \{1_{A(x) \geq \tth} O(x) : x \in \SSe\}\\
Z_2(\tth) &\coloneqq \{1_{A(x) < \tth} O(x) : x \in \SSe\}.
\end{align*}
Note that $\frac{\hat{\mu}_{Z_1(\tth)}}{\hat{\mu}_{Z_1(\tth)} + \hat{\mu}_{Z_2(\tth)}} = \text{Recall}_{\SSe}(\tth)$,
which increases with $\hat{\mu}_{Z_1(\tth)}$ and decreases with $\hat{\mu}_{Z_2(\tth)}$.
Thus, if we let 
\begin{align*}
\tm^{*} = \frac{\text{UB}(\mu_{Z_1(\tth_o)}, \sigma_{Z_1(\tth_o)}, s, \frac{\delta}{2})}{
	\text{UB}(\mu_{Z_1(\tth_o)}, \sigma_{Z_1(\tth_o)}, s, \frac{\delta}{2}) +
	\text{LB}(\mu_{Z_2(\tth_o)}, \sigma_{Z_2(\tth_o)}, s, \frac{\delta}{2})
}
\end{align*}
then asymptotically as $s\to \infty$ Lemma~\ref{lem:clt} ensures
$
\text{Recall}_{\SSe}(\tth_o) = \frac{\hat{\mu}_{Z_1(\tth_o)}}{\hat{\mu}_{Z_1(\tth_o)} + \hat{\mu}_{Z_2(\tth_o)}} \leq \tm^{*}
$
with probability $1-\delta$.
$\tm^{*}$ is not computable from our sample so we use plug-in estimates for $\tth_o$,
$\mu$, and $\sigma$ to estimate a $\tm' \rightarrow \tm^{*}$ as $s\rightarrow \infty$.

\subsubsection{Precision Target}
For precision target queries, we want to estimate a threshold $\tth$ such that 
$\text{Precision}_{\SD}(\tth) \geq \tm$
with high probability.
To maximize result quality (i.e., maximize recall), we would further like to
make $\tth$ as small as possible.

Unlike for recall target queries, there is no monotonic relationship between $\text{Precision}_{\SD}(\tth)$
and $\tth$: 
 $\text{Precision}_{\SD}(\tth_1)$ may be greater than $\text{Precision}_{\SD}(\tth_2)$
even if $\tth_1 < \tth_2$.
Thus, for PT queries we calculate lower bounds on the precision provided by a large set of candidate thresholds $\tth$, and return the smallest candidate threshold that provides
results with precision above the target.

\begin{algorithm}[t!]
\begin{algorithmic}
\Function{$\tth_{\text{\AlgBR}}$}{$\SD$}
  \State $\SSe \gets \text{UniformSample}(\SD, s)$
  \State $\hat{\tth}_o \gets \max \{\tth : \text{Recall}_{\SSe}(\tth) \geq \tm \}$
  \State $Z_1 \gets \{1_{A(x) \geq \hat{\tth}_o} O(x) : x \in \SSe\}$
  \State $Z_2 \gets \{1_{A(x) < \hat{\tth}_o} O(x) : x \in \SSe\}$
  \State $\tm' \gets \frac{
	\text{UB}(\hat{\mu}_{z_1}, \hat{\sigma}_{z_1}, s, \delta/2)
	}{
	\text{UB}(\hat{\mu}_{z_1}, \hat{\sigma}_{z_1}, s, \delta/2) + \text{LB}(\hat{\mu}_{z_2}, \hat{\sigma}_{z_2}, s, \delta/2)
	}$
  \State $\tth' \gets \max \{\tth : \text{Recall}_{\SSe}(\tth) \geq \tm' \}$
  \State \textbf{return} $\tth'$
\EndFunction
\end{algorithmic}
\caption{Uniform threshold estimation (RT)}
\label{alg:thresh-rt}
\end{algorithm}

\begin{algorithm}[t!]
\begin{algorithmic}
\State $m \gets 100$ \Comment{Minimum step size}
\Function{$\tth_{\AlgBP}$}{$\SD$}
  \State $\SSe \gets \text{UniformSample}(\SD, s)$
  \State $A_{\SSe} \gets \text{Sort}(\{A(x) : x \in \SSe\})$
  \State $M \gets \lceil s / m \rceil$

  \State $\text{Candidates} \gets \{\}$
  \For{$i \gets m,2m,\dots,s$} 
  	\State $\tth \gets A_{\SSe}[i]$
  	\State $Z \gets \{O(x): x\in \SSe \wedge A(x) \geq \tth\}$
  	\State $p_l \gets \text{LB}(\hat{\mu}_{Z}, \hat{\sigma}_{Z}, |Z|, \delta/M)$
  	\Comment{Precision Bound}
  	\If{$p_l > \tm$}
  		\State $\text{Candidates} \gets \text{Candidates} \cup \{\tth\}$
  	\EndIf
  \EndFor
  \State \textbf{return} $\min_{\tth} \text{Candidates}$
\EndFunction
\end{algorithmic}
\caption{Uniform threshold estimation (PT)}
\label{alg:thresh-pt}
\end{algorithm}

We provide pseudocode for \AlgBR which uses confidence intervals over a uniform sample (Algorithm~\ref{alg:thresh-pt}).
Since the procedure uses Lemma~\ref{lem:clt} $M$ times by union bound we need each usage 
to hold with probability $1-\delta/M$ for the final
returned threshold to be valid with probability $1-\delta$.

\minihead{Validity justification}
Let 
$$Z(\tth) = \{O(x): x\in \SSe \wedge A(x) \geq \tth\},$$
then
$\hat{\mu}_{Z(\tth)} = \text{Precision}_{\SSe}(\tth)$ and
$\mu_{Z(\tth)} = \text{Precision}_{\SD}(\tth).$
Asymptotically by Lemma~\ref{lem:clt}, with probability $1-\delta/M$
$$
\text{LB}(\hat{\mu}_{Z(\tth)}, \sigma_{Z(\tth)}, |Z(\tth)|, \delta/M)
\leq \mu_{Z(\tth)}.
$$
By the union bound, as long as each $\tth$ in the $\text{Candidate}$ set
has $\text{LB}(\hat{\mu}_{Z(\tth)}, \sigma_{Z(\tth)}, |Z(\tth)|, \delta/M) > \tm$,
the precision for each of the candidates over the dataset also
exceeds $\tm$.
Since we do not know $\sigma$, in Algorithm~\ref{alg:thresh-pt} we use
sample plug-in estimates for $\sigma_{Z(\tth)}$. Alternatively one could
use a t-test (both are asymptotically valid).

\subsection{Importance Sampling}
\label{sec:importance}
The \uniform-\ci routines for estimating $\tth$
in Algorithms~\ref{alg:thresh-rt} and \ref{alg:thresh-pt}
provide valid results with probability $1-\delta$. 
However if the
random sample chosen for oracle labeling $\SSe$ is uninformative,
the confidence bounds we use will be wide and the threshold
estimation routines will return results that have lower quality in
order to provide valid results.
Thus, we explain how \sn uses importance sampling
to select a set of points that improve upon uniform sampling.
We refer to these more efficient routines as \imp-\ci estimators.

Importance sampling chooses records $x$ with replacement from the dataset
$\SD$ with weighted probabilities $w(x)$ as opposed to uniformly
with base probability $u(x)$.
One can compute the expected value of a quantity $f(x)$ 
with reduced variance by then sampling according
to $w$ rather than $u$ and using the reweighting identity:
\begin{align}
\ex_{x \sim u}[f(x)] = \ex_{x \sim w}\left[f(x) \frac{u(x)}{w(x)}\right].
\end{align}

\vspace{3em}

Abbreviating the reweighting factor as $m(x) \coloneqq u(x)/w(x)$,
we can then define reweighted estimates for recall and precision
on a weighted sample $\SSe_w$:
\begin{align}
\text{Recall}_{\SSe_w}(\tth) &\coloneqq \frac{\sum_{x \in \SSe} 1_{A(x) \geq \tth} O(x) m(x)}{\sum_{x \in \SSe_w} O(x) m(x)} \\
\text{Precision}_{\SSe_w}(\tth) &\coloneqq \frac{\sum_{x \in \SSe} 1_{A(x) \geq \tth} O(x) m(x)}{\sum_{x \in \SSe_w} m(x)} 
\end{align}
If we can reduce the variance of these estimates, we can use the tighter bounds to
improve the quality of the results at a given recall or precision target.

\sloppypar{
The optimal choice of $w(x)$ for the standard importance sampling setting is $w(x)
\propto f(x)u(x)$~\cite{wasserman2013all}.
However, this assumes $f(x)$ is a known function. 
In our setting, we want $f(x) = 1_{A(x) \geq \tth} O(x)$ which is both stochastic and a priori
unknown. This prevents us from directly applying traditional importance sampling weights based on $f(x)$.
Instead, we can use the proxy $A(x)$ to define sampling weights.
}

Our approach solves for the optimal sample weights 
for proxies that are highly correlated with the oracle, i.e. \emph{calibrated}
with $A(x) = \Pr_{x \sim u}[O(x)=1 | A(x)]$.
In practice this will not hold exactly, but as long as the proxy scores are
approximately proportional to the probability we can use them to 
derive useful sample weights.
We show in Section~\ref{sec:stat_efficiency} that the optimal
weights which minimize the variance are proportional to $\sqrt{A(x) 1_{A(x) \geq \tth}}u(x)$.
To guard against situations where the proxy could be inaccurate, we
defensively mix a uniform distribution with these optimal weights in our algorithms
\cite{owen2000safeis}.

Note that the validity of our results does not depend on the proxy
being calibrated, but this importance sampling scheme allows us to
obtain lower variance threshold estimates and thus more efficient
query results when the proxy is close to calibrated.

\vspace{0.3em}

\minihead{Recall target}
For recall target queries, we extend Algorithm~\ref{alg:thresh-rt}
to use weighted samples according to Theorem~\ref{thm:opt_imp_sample}.
We use the weights to optimize the variance of $E[O(x)]$ as a proxy for
reducing the variance of $E[1_{A(x) \geq \tth_o} O(x)]$ and $E[1_{A(x) < \tth_o} O(x)]$.
We present this weighted method, \AlgCR, in Algorithm~\ref{alg:thresh-final-rt}.
The justification for high probability validity is the same as before.

\begin{algorithm}[t!]
\begin{algorithmic}
\Function{$\tth_{\text{\AlgCR}}$}{$\SD$}
  \State $\vec{w} \gets \{\sqrt{A(x)} : x \in \SD\}$
  \State $\vec{w} \gets .9\cdot \vec{w} / \|\vec{w}\|_1 + .1\cdot \vec{1}/|\SD|$ \Comment{Defensive Mixing}
  \State $\SSe \gets \text{WeightedSample}(\SD, \vec{w}, s)$
  \State $m(x) \gets \frac{1/|\SD|}{w(x)}$
  \State $\tth_o \gets \max \{\tth : \text{Recall}_{\SSe_w}(\tth) \geq \tm \}$
  \State $\hat{z}_1 \gets \{1_{A(x) \geq \tth_o} O(x) m(x) : x \in \SSe\}$
  \State $\hat{z}_2 \gets \{1_{A(x) < \tth_o} O(x) m(x) : x \in \SSe\}$
  \State $\tm' \gets \frac{
	\text{UB}(\hat{\mu}_{z_1}, \hat{\sigma}_{z_1}, s, \delta/2)
	}{
	\text{UB}(\hat{\mu}_{z_1}, \hat{\sigma}_{z_1}, s, \delta/2) + \text{LB}(\hat{\mu}_{z_2}, \hat{\sigma}_{z_2}, s, \delta/2)
	}$
  \State $\tth' \gets \max \{\tth : \text{Recall}_{\SSe_w}(\tth) \geq \tm' \}$

  \State \textbf{return} $\tth'$
\EndFunction
\end{algorithmic}
\caption{Importance threshold estimation (RT)}
\label{alg:thresh-final-rt}
\end{algorithm}

\vspace{0.3em}

\minihead{Precision target}
For PT queries we can combine Theorem~\ref{thm:opt_imp_sample} with an additional
observation: if we know there are at most $n_{\text{match}}$ positive matching
records in $\SD$, then there is no need to consider thresholds lower than the
$n_{\text{match}} / \tm$-th highest proxy score in $\SD$, since any lower
thresholds cannot achieve a precision of $\tm$.
\sn thus devotes half of the oracle sample budget to estimating the upper bound
$n_{\text{match}}$ and the remaining half for running
a weighted version of Algorithm~\ref{alg:thresh-pt} on candidate thresholds.
We present this two-stage weighted sampling algorithm, 
\AlgCP, in Algorithm~\ref{alg:thresh-final-pt}.

\begin{algorithm}[t!]
\begin{algorithmic}
\State $m \gets 100$ \Comment{Minimum step size}

\Function{$\tth_{\text{\AlgCP}}$}{$\SD$}
  \State $\vec{w} \gets \{\sqrt{A(x)} : x \in \SD\}$
  \State $\vec{w} \gets .9\cdot \vec{w} / \|\vec{w}\|_1 + .1\cdot \vec{1}/|\SD|$ \Comment{Defensive Mixing}
  
  \State $\SSe_0 \gets \text{WeightedSample}(\SD, w, s/2)$
  \Comment{Stage 1}
  \State $m(x) \gets \frac{1/|\SD|}{w(x)}$
  \State $Z \gets \{O(x) m(x) : x \in \SSe_0\}$
  \State $n_{\text{match}} \gets |\SD| \cdot \text{UB}(\hat{\mu}_{Z}, \hat{\sigma}_{Z}, s/2, \delta/2)$

  \State $A \gets \text{SortDescending}(\{A(x) : x \in \SD\})$
  \State $\SD' \gets \{x : A(x) \geq A[n_{\text{match}}/\tm]\}$

  \State $\SSe_1 \gets \text{WeightedSample}(\SD', w, s/2)$
  \Comment{Stage 2}
  \State $A_{\SSe_1} = A \cap \SSe_1$
  
  \State $M \gets \lceil s / m \rceil$
  \State $\text{Candidates} \gets \{\}$
  \For{$i \gets m,2m,\dots,s$} 
  	\State $\tth \gets A_{\SSe_1}[i]$
  	\State $Z \gets \{O(x): x\in \SSe_1 \wedge A(x) \geq \tth\}$
  	\State $p_l \gets \text{LB}(\hat{\mu}_{Z}, \hat{\sigma}_{Z}, |Z|, \delta/(2M))$
  	\Comment{Precision Bound}
  	\If{$p_l > \tm$}
  		\State $\text{Candidates} \gets \text{Candidates} \cup \{\tth\}$
  	\EndIf
  \EndFor
  \State \textbf{return} $\min_{\tth} \text{Candidates}$
\EndFunction
\end{algorithmic}
\caption{Importance threshold estimation (PT)}
\label{alg:thresh-final-pt}
\end{algorithm}

We set the failure probability of each stage to $\delta/2$ which guarantees
the overall failure probability of the algorithm via the union bound.
The remaining arguments for high probability validity follows the
argument for the unweighted algorithm.

\vspace{0.3em}

\subsubsection{Statistical Efficiency}
\label{sec:stat_efficiency}

\minihead{Algorithm}
Theorem~\ref{thm:opt_imp_sample}  formally states the optimal sampling weights 
used by our importance sampling $\tth$ estimation routines.
The proof is deferred to Section~\ref{sec:thm_imp_proof}.
\begin{theorem}
For an importance sampling routine estimating $\ex_{x \sim u}[f(x)]$,
when $f(x) = \{0,1\}$, $a(x)$ is a calibrated proxy
$\Pr_{x \sim u}[f(x)=1| a(x)] = a(x),$
and we sample knowing $a(x),u(x)$, but not $f(x)$,
then importance sampling with $w(x) \propto \sqrt{a(x)}u(x)$
minimizes the variance of the reweighted estimator.
\label{thm:opt_imp_sample}
\end{theorem}

We apply can this to our algorithms using 
$f(x) = O(x) \cdot 1_{A(x) \geq \tth}$ and
$a(x) = A(x) \cdot 1_{A(x) \geq \tth}$.
To illustrate the impact of these weights, we can quantify the maximum
improvement in variance they provide.
Compared with uniform sampling or sampling proportional to $a(x)$, 
these weights provide a variance reduction
of at least $\Delta_{v} = \var_{x \sim u}[\sqrt{a(x)}]$, which is significant when
the proxy confidences are concentrated near $0$ and $1$, while the
differences vanish when there is little variation in the proxy scores.
For more details see Section~\ref{sec:varcompare}.

\vspace{0.3em}

\minihead{Intuition}
\colora{
In standard importance sampling, the variance minimizing weights are
proportional to the function values. However, in our setting, we only have access
to probabilities (i.e., $A(x)$) for the function we wish to
compute expectations over (i.e., $O(x)$). 
Since $O(x)$ is a randomized realization of $A(x)$,
up-weighing $x$ proportionally to $A(x)$ results in ``overconfident'' sampling.
Thus, the square root weights effectively down-weights the confidence that
$A(x)$ accurately reflects $O(x)$. We show in
Section~\ref{sec:eval-sensitivity} the effect
of the exponent in weighing $A(x)$ on the sample efficiency.
}

\begin{table*}[th!]
\centering
\caption{Summary of datasets, oracle models, proxy models, and true positive
rates. We use both synthetic and real datasets that vary in true positive rate
and type of proxy/oracle models.}
\label{table:datasets}
\vspace{-0.3em}
\begin{tabularx}{\textwidth}{llllX}
  Dataset  & Oracle & Proxy & TPR & Task description \\ \hline
  ImageNet            & Human labels & ResNet-50 & 0.1\% & Finding hummingbirds in the ImageNet validation set \\
  \texttt{night-street} & Mask R-CNN & ResNet-50 & 4\%   & Finding cars in the \texttt{night-street} video \\
  OntoNotes           & Human labels & LSTM      & 2.5\% & Finding city relationships \\
  TACRED              & Human labels & SpanBERT  & 2.4\% & Finding employees relationships \\
  $\mathrm{Beta}(0.01, 1)$ & True values & Probabilities & 0.5\% & $A(x) = \mathrm{Beta}(0.01, 1)$ and $O(x) = \mathrm{Bernoulli}(A(x))$ \\
  $\mathrm{Beta}(0.01, 2)$ & True values & Probabilities & 1\%   & We use the same procedure as directly above but with $\mathrm{Beta}(0.01, 2)$
\end{tabularx}
\vspace{-0.3em}
\end{table*}

\vspace{1em}

\section{Evaluation}
\label{sec:eval}

We evaluate our algorithms on six real-world and synthetic datasets. We
describe the experimental setup, demonstrate that naive algorithms fail to
respect failure probabilities, demonstrate that our algorithms outperform
uniform sampling (as used by prior work), and that our algorithms are robust to
proxy choices.

\subsection{Experimental Setup}

\subsubsection{Metrics}
Following the query definitions in Section~\ref{sec:queries}, we are interested in two primary evaluation metrics:
\begin{enumerate}
\item We measure the empirical failure rate of the different algorithms: the rate at which
they do not achieve a target recall or precision.

\item We measure the quality of query results using achieved precision when there
is a minimum target recall, and achieved recall when there is a minimum target precision.
\end{enumerate}

\subsubsection{Methods Evaluated}
In our evaluation we compare methods that all select records based on
a proxy threshold as in Algorithm~\ref{alg:query-process}.
The methods differ in their sampling routine and routine for estimating
the proxy threshold $\tth$ as described in Section~\ref{sec:alg}.
Note that \noscope and probabilistic predicates correspond to the baseline algorithms
\AlgAR and \AlgAP with no guarantees.
We can extend these algorithms to provide probabilistic guarantees in the
\AlgBR and \AlgBP algorithms.
Finally, our system \sn uses the \AlgCR and \AlgCP algorithms which introduce
importance sampling.\footnote{Code for our algorithms is available at \url{https://github.com/stanford-futuredata/supg}.}

Many systems additionally compare against full scans. However, this baseline
always requires executing the oracle model on the entire dataset $\universe$,
requiring $|\universe|$ oracle model invocations. 
On large datasets, this approach was infeasible for our collaborators and
industry contacts, so we exclude this baseline from comparison.

\subsubsection{Datasets and Proxy Models}
We show a summary of datasets used in Table~\ref{table:datasets}.

\minihead{Beta (synthetic)}
We construct synthetic datasets using proxy scores $A(x)$ drawn
from a $\textrm{Beta}(\alpha, \beta)$ distribution, allowing us to
vary the relationship between the proxy model and oracle labels. 
We assign ground truth oracle labels as independent Bernoulli trials
based on the proxy score probability.
These synthetic datasets have $10^6$ records and
we use two pairs of $(\alpha, \beta)$: (0.01, 1) and (0.01, 2).

\minihead{ImageNet and \texttt{night-street} (image)}
We use two real-world image datasets to evaluate \sn. First, we use the ImageNet
validation dataset~\cite{russakovsky2015imagenet} and select instances of
hummingbirds. There are 50 instances of hummingbirds out of 50,000 images or an
occurrence rate of 0.1\%. The oracle model is human labeling. Second, we use the
commonly used \texttt{night-street} video~\cite{kang2017noscope,
kang2019blazeit, xu2019vstore, canel2019scaling} and select cars from the video.
The oracle model is an expensive, state-of-the-art object detection
method~\cite{he2017mask}. We resample the positive instances of cars to set the
true positive rate to 4\% to better model real-world scenarios where matches
are rare.
Note that our algorithms typically perform better under higher class imbalance.

The proxy model for both datasets is a ResNet-50~\cite{he2016deep}, which is
significantly cheaper than the oracle model.

\minihead{OntoNotes and TACRED (text)}
\colora{
We use two real-world text datasets (OntoNotes~\cite{hovy2006ontonotes} with
fine-grained entities~\cite{choi2018ultra} and TACRED~\cite{zhang2017tacred}) to
evaluate \sn. The task for both datasets is relation extraction, in which the
goal is to extract semantic relationships from text, e.g., ``organization'' and
``founded by.'' We searched for city and employees relationships for OntoNotes
and TACRED respectively. The oracle model is human labeling for both datasets.

The proxy model for OntoNotes is a baseline provided by with the fine-grained
entities~\cite{choi2018ultra}. The proxy model for TACRED is the
state-of-the-art SpanBERT model~\cite{joshi2020spanbert}. We choose different
models to demonstrate that \sn is agnostic to proxy model choice.
}

\subsection{Baseline~Methods~Fail~to\\Achieve~Guarantees}
\label{sec:eval-guarantees}
We demonstrate that baseline methods fail to achieve guarantees on failure
probability. First, we show that \uniform-\noci (i.e., uniform sampling from the
universe and choosing the empirical cutoff, Section~\ref{sec:alg}) fails. Note
that \uniform-\noci is used by prior work. Second, we show that using
\uniform-\noci on other data, as other systems do, also fails to achieve the
failure probabilities.

\minihead{\uniform-\noci fails}
To demonstrate that \uniform-\noci fails to achieve the failure probability, we
show the distribution of precisions and recalls under 100 trials of this
algorithm and \sn's optimized importance sampling algorithm. For \sn, we set
$\delta = 0.05$. We targeted a precision and recall of 90\% for both methods.

\begin{figure}[t!]
  \includegraphics[width=\columnwidth]{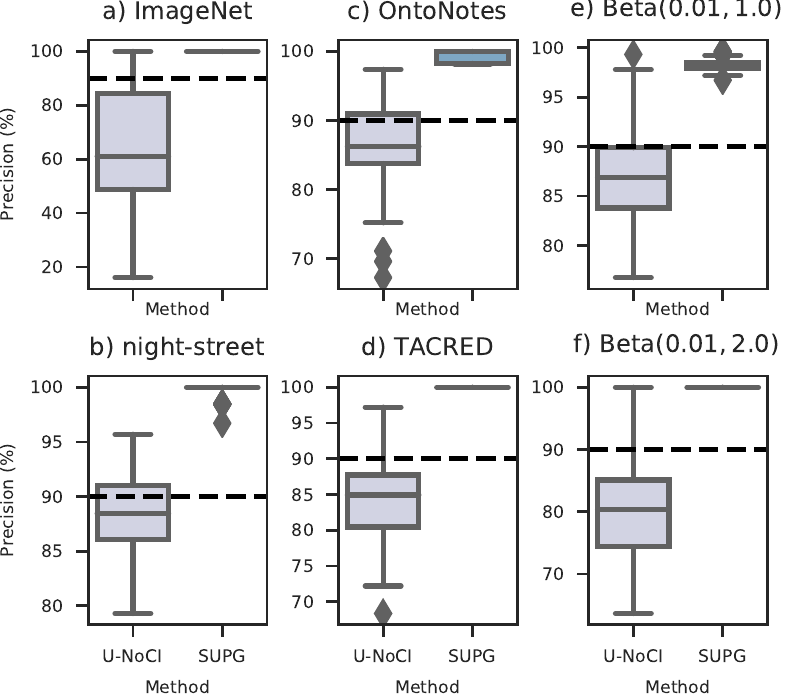}
  \vspace{-0.5em}
  \caption{Precision of 100 trials of \uniform-\noci and \sn's importance
  sampling algorithm with a precision target of 90\%. We show a box plot, in
  which the box 25th, 50th, and 75th quantiles; the minimum and maximum
  excluding outliers are the ``whiskers.'' As shown, \uniform-\noci can fail up
  to 75\% of the time. Furthermore, it can return precisions as low as 20\%.}
  \label{fig:naive-precision}
  \vspace{-0.5em}
\end{figure}

\begin{figure}[t!]
  \includegraphics[width=\columnwidth]{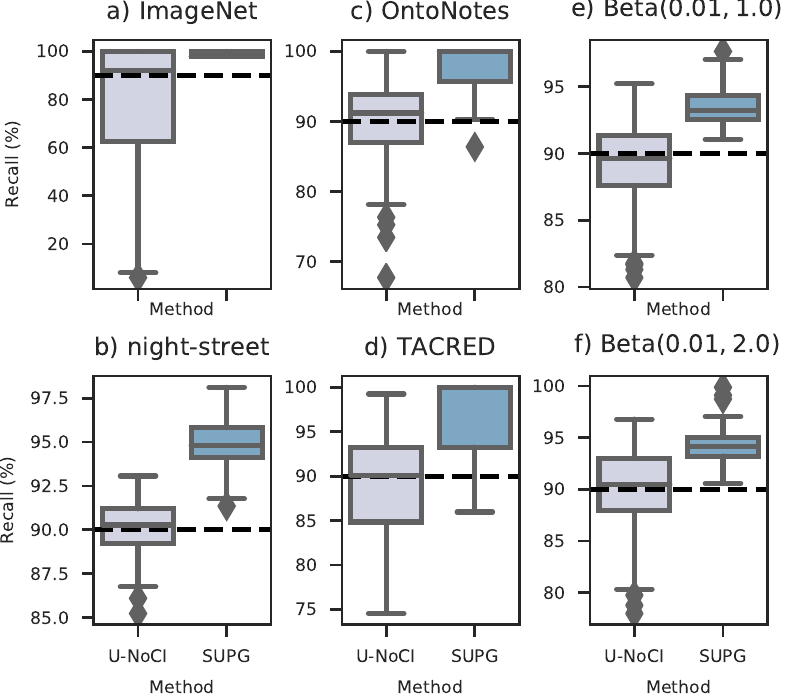}
  \vspace{-0.5em}
  \caption{Recall of 100 trials of \uniform-\noci and \sn's importance sampling
  algorithm with a recall target of 90\%. As shown, \uniform-\noci can fail up
  to 50\% of the time and even catastrophically fail on ImageNet, returning a
  recall of as low at 20\%.}
  \label{fig:naive-recall}
  \vspace{-0.5em}
\end{figure}

As shown in Figures \ref{fig:naive-precision} and \ref{fig:naive-recall},
\uniform-\noci can fail as much as 75\% of the time. Furthermore, \uniform-\noci
can catastrophically fail, returning recalls of under 20\% when 90\% was
requested. In contrast, \sn's algorithms respect the recall targets within the
given $\delta$.

\begin{table}[t!]
\centering
\caption{Summary of distributionally shifted datasets. These shifts are natural
(weather related, different day of video) and synthetic.}
\label{table:dataset-drift}
\vspace{-0.5em}
\begin{tabularx}{\columnwidth}{llX}
  Dataset  & Shifted dataset & Description \\ \hline
  ImageNet & ImageNet-C, Fog & ImageNet with fog \\
  \texttt{night-street} & Day 2 & Different days \\
  $\mathrm{Beta}(0.01, 1)$ & $\mathrm{Beta}(0.01, 2)$ & Shifted $\beta$ parameter
\end{tabularx}
\vspace{-0.3em}
\end{table}

\minihead{\uniform-\noci fails under model drift}
We further show that \uniform-\noci on different data distributions also fails
to achieve the failure probability. This procedure is used by existing systems
such as \noscope and probabilistic predicates on a given set of data; the cutoff
is then used on other data. These systems assume the data distribution is fixed,
a known limitation. 

To evaluate the effect of model drift, we allow the \uniform-\noci to choose
a proxy threshold using
oracle labels on \emph{the entire training dataset} and then
perform selection on test datasets with distributional shift. 
We compare this with applying the \sn algorithms using a limited
number of oracle labels from the shifted test set as usual.
We summarize the shifted datasets in Table~\ref{table:dataset-drift}. We
use naturally occurring instances of drift (obscuration by
fog~\cite{hendrycks2018benchmarking}, different day of video) and synthetic
drift (change of $\mathrm{Beta}$ parameters).

As shown in Table~\ref{table:drift-guarantees}, baseline methods
that do not use labels from the shifted dataset fail
to achieve the target in all settings, even under mild conditions such as
different days of a video. In fact, using the empirical cutoff
in \uniform-\noci can result in
achieved targets as much as 41\% lower. In contrast, our algorithms will always
respect the failure probability despite model drift, addressing a limitation in
prior work~\cite{kang2017noscope, kang2019blazeit, lu2018accelerating,
anderson2018predicate, hsieh2018focus}.

\begin{table}[t!]
\centering
\caption{Achieved accuracy of queries when using the empirical cutoff method and
\sn on data with distributional shift. We show the average of 100 runs for \sn.
All methods targeted a success rate of 95\%. As shown, the naive algorithm
\emph{deterministically fails} to achieve the targets, i.e., has a failure rate
of 100\%.}
\label{table:drift-guarantees}
\vspace{-0.5em}
\begin{tabular}{lllll}
             & Query     &        & Naive    & \sn \\
  Dataset    & type      & Target & accuracy & accuracy \\ \hline
  ImageNet-C & Precision & 95\% & 77\% & 100\% \\
  ImageNet-C & Recall    & 95\% & 54\% & 100\% \\
  \texttt{night-street} & Precision & 95\% & 89\% & 97\% \\
  \texttt{night-street} & Recall    & 95\% & 89\% & 96\% \\
  Beta & Precision & 95\% & 89\% & 100\% \\
  Beta & Recall    & 95\% & 90\% & 98\%
\end{tabular}
\vspace{-0.6em}
\end{table}

\subsection{\sn Outperforms Uniform Sampling}
\label{sec:eval_fu}
We show that \sn's novel algorithms for selection outperforms \uniform-\ci
(i.e., uniform sampling with guarantees) in both the precision target and recall
target settings. Recall that the goal is to maximize or minimize the size of the
returned set in the precision target and recall target settings, respectively.

\minihead{Precision target setting}
For the datasets and models described in Table~\ref{table:datasets}, we executed
\uniform-\ci, one-stage importance sampling, and two-stage importance
sampling for the precision target setting. We used a budget of 1,000 oracle
queries for ImageNet and 10,000 for \texttt{night-street} and the synthetic
dataset. We targeted precisions of 0.75, 0.8, 0.9, 0.95, and 0.99.

\begin{figure}[t!]
  \includegraphics[width=\columnwidth]{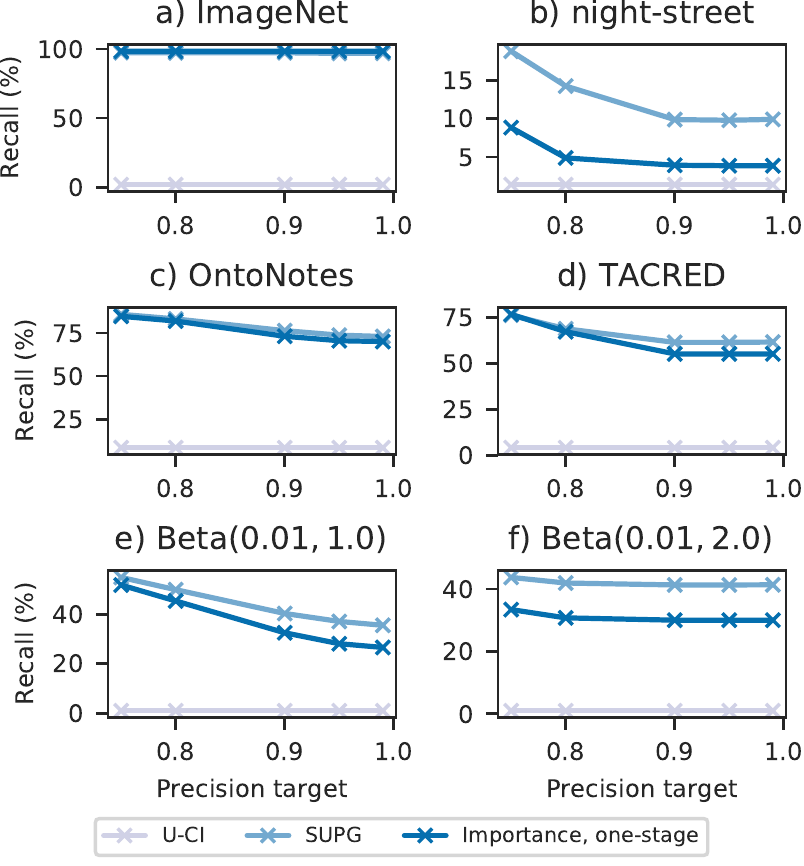}
  \caption{Targeted precision vs achieved recall. As shown,
  both importance sampling methods outperform \uniform-\ci in all cases.
  Two-stage importance sampling outperforms all methods and matches the
  one-stage importance sampling for ImageNet.
  }
  \label{fig:fu-precision}
\end{figure}

We show the achieved precision and recall for the various methods in
Figure~\ref{fig:fu-precision}. As shown, the importance sampling method
outperforms \uniform-\ci in all cases. Furthermore, the two-stage algorithm
outperforms or matches the one-stage algorithm in all cases except ImageNet.
While the specific recalls that are achieved vary per dataset, this is largely
due to the performance of the proxy model.

We note that the ImageNet dataset and proxy model are especially favorable to
\sn's importance sampling algorithms. This dataset has a true positive rate of
0.1\% and a highly calibrated proxy. A low true positive rate will result in
uniform sampling drawing few positives. In contrast, a highly calibrated proxy
will result in many positive draws for importance sampling.

\minihead{Recall target setting}
For the datasets and models in Table~\ref{table:datasets}, we executed
\uniform-\ci, standard importance sampling with linear weights $\propto A(x)$
(Importance, prop), and the \sn methods that use sqrt weights. We used the same
budgets as in the precision target setting. We targeted recalls of 0.5, 0.6,
0.7, 0.75, 0.8, 0.9, and 0.95.

\begin{figure}[t!]
  \includegraphics[width=\columnwidth]{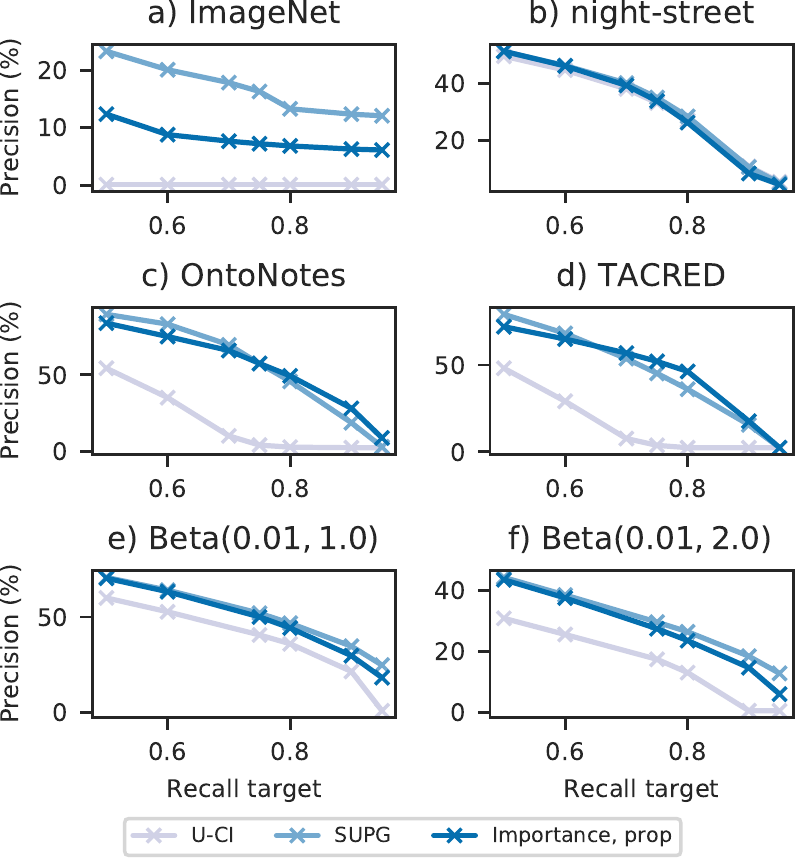}
  \caption{Targeted recall vs precision
  of the returned set. Up and to the right indicates higher performance.
  Importance sampling outperforms or matches \uniform-\ci in all cases. Our
  sqrt scaling outperforms proportional scaling for importance sampling in all
  cases, except for high recall settings.
  }
  \label{fig:fu-recall}
\end{figure}

We show the achieved recall and the returned set size for the various methods in
Figure~\ref{fig:fu-recall}. As shown, the importance sampling method outperforms
\uniform-\ci in all cases. Furthermore, using $\sqrt{A(x)}$ weights
outperforms using linear weights in all cases.

\subsection{Sensitivity Analysis}
\label{sec:eval-sensitivity}

We analyze how sensitive our novel algorithms are to: 1) the performance of the
proxy model, 2) the class imbalance ratio,
\colora{
and 3) the parameters in our algorithms.
}

\minihead{Sensitivity to proxy model}
We analyze the sensitivity to the proxy model in two ways: 1) we add noise to
the generating distribution for the synthetic dataset and 2) we vary the
parameters $\alpha$ and $\beta$ to vary the sharpness of the generating
distribution.

\begin{figure}[t!]
  \includegraphics[width=\columnwidth]{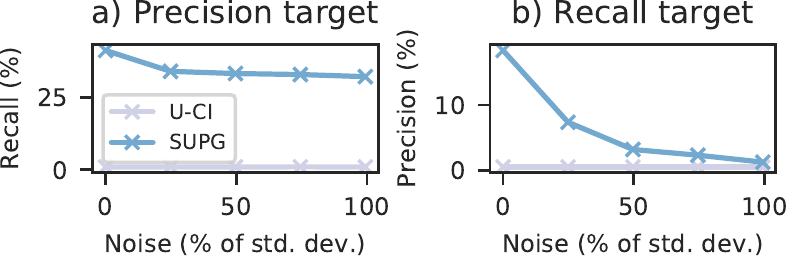}
  \vspace{-0.5em}
  \caption{Noise level vs recall/precision for the precision/recall target
  settings, respectively. The noise level is given as a percent of the standard
  deviation of the original probabilities. As shown, \sn outperforms uniform
  sampling at all noise levels, even up to 100\% noise.}
  \label{fig:synthetic-noise}
  \vspace{-0.5em}
\end{figure}

First, we generate oracle values from $\mathrm{Beta}(0.01, 2)$. After oracle
values are generated, we add Gaussian noise to the proxy scores and clip them to
$[0, 1]$. We add Gaussian noise with standard deviations of 0.01, 0.02, 0.03,
and 0.04, which corresponds to 25\%, 50\%, 75\%, and 100\% of the standard
deviation of the original probabilities. We targeted a precision and recall of
95\% and 90\% respectively.
We show results in Figure~\ref{fig:synthetic-noise}. As shown, while the
performance of our algorithms degrades with higher noise, importance sampling
still outperforms uniform sampling at all noise levels. Furthermore, \sn's
algorithms degrade gracefully with higher noise, especially in the precision
target setting.

Second, we vary $\alpha$ and $\beta$ to vary the sharpness. We find that varying
$\beta$ as described below (class imbalance) also changes the sharpness of the
distribution, as measured by the standard deviation of the probabilities. As the
results are the same, we defer the discussion to below. We note that \sn
outperforms uniform sampling in all cases and degrades gracefully as the
sharpness of the proxy model decreases.

\minihead{Sensitivity to class imbalance}
We analyze the sensitivity of our algorithms to class imbalance by varying
$\alpha$ and $\beta$. We fix $\alpha$ at 0.01 and set $\beta \in \{ 0.125, 0.25,
0.5, 1.0, 2.0\}$.

\begin{figure}[t!]
  \includegraphics[width=\columnwidth]{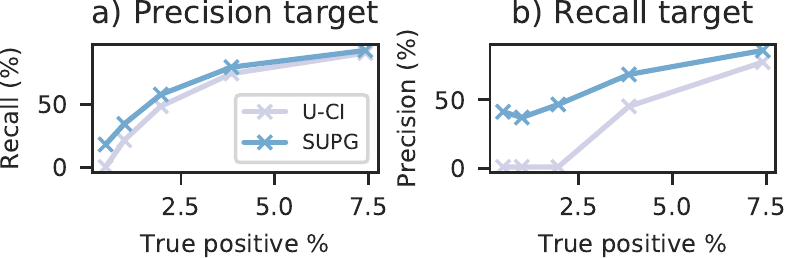}
  \vspace{-0.5em}
  \caption{True positive rate vs recall/precision for the precision/recall
  target settings, respectively. As shown, \sn outperforms uniform sampling in
  all scenarios, even as the true positive rate is as high as 7\%. \sn
  especially outperforms uniform sampling at low true positive rates,
  outperforming by as much as 47$\times$.}
  \label{fig:class-imbalance}
  \vspace{-0.5em}
\end{figure}

We show results for varying these values in Figure~\ref{fig:class-imbalance}. As
shown, our algorithms outperform uniform sampling more as the class imbalance is
higher. High class imbalance is common in practice, so we optimize our
algorithms for such cases. For these cases, \sn outperforms by as much as
47$\times$. As the data becomes more balanced, our algorithms outperform uniform
sampling less, but still outperforms uniform sampling.

\begin{figure}[t!]
  \includegraphics[width=\columnwidth]{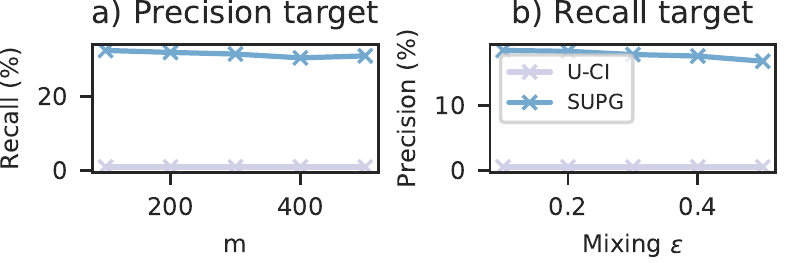}
  \caption{\colora{Effect of parameter settings on algorithm performance. As
  shown, \sn performs well across a range of parameter settings, indicating that
  parameters are not difficult to set.}}
  \label{fig:vary-params}
\end{figure}

\minihead{Sensitivity to parameters}
\colora{
We analyze the sensitivity of our algorithms to parameter settings ($m$ in
Algorithm~\ref{alg:thresh-final-pt} and the defensive mixing ratio in
Algorithm~\ref{alg:thresh-final-rt}). We vary $m$ from 100 to 500 in increments
of 100 and the mixing ratio from 0.1 to 0.5 in increments of 0.1 for
$\mathrm{Beta}(0.01, 2)$. As shown in Figure~\ref{fig:vary-params}, \sn performs
well across a range of parameters. We note that some defensive mixing is
required to avoid catastrophic failing, but these results indicate that our
parameters are not difficult to set by choosing any value away from 0 and 1.

}

\begin{figure}[t!]
  \includegraphics[width=\columnwidth]{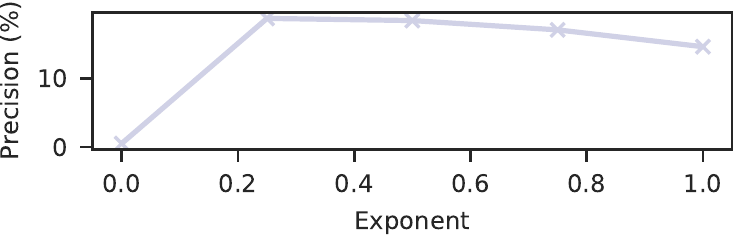}
  \vspace{-0.5em}
  \caption{\colora{Exponent in importance sampling weights vs precision for the
  recall target setting. As shown, exponents closer to 0.5 perform better.}}
  \label{fig:vary-exp}
\end{figure}

\minihead{Sensitivity to exponent}
\colora{
We analyze the sensitivity to the importance weight exponent by varying it from
0 to 1 for the recall target for
$\mathrm{Beta}(0.01, 2)$. As shown in Figure~\ref{fig:vary-exp}, exponents 
corresponding to uniform ($0$) and proportional ($1$) sampling do not perform well. 
Square root weighting is close to optimal. We
note that while our proof shows square root weighting is optimal for estimating
counts, optimal end-to-end performance may require slightly different weights.
Nonetheless, it outperforms exponents of 0 and 1 and performs well in practice.

}

\begin{figure}[t!]
  \includegraphics[width=\columnwidth]{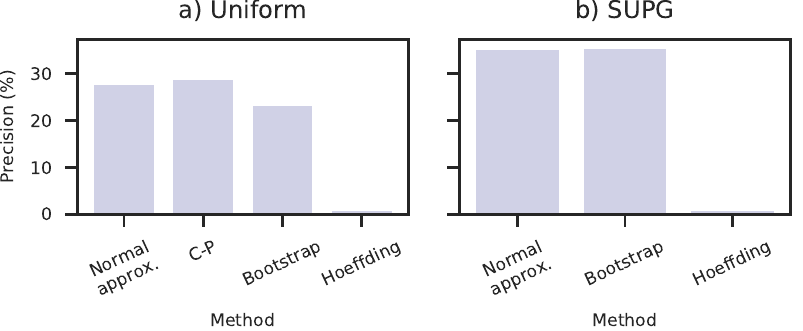}
  \vspace{-0.5em}
  \caption{\colora{Performance of \AlgBR using various confidence interval
  methods. As shown, the normal approximation matches or outperforms
  alternatives, within error margins.}}
  \label{fig:ci}
  \vspace{-0.5em}
\end{figure}

\minihead{Sensitivity to confidence interval method}
\colora{
We analyze how the confidence interval method affects performance. We consider
the normal approximation~\cite{bentkus1996berry}, Hoeffding's
inequality~\cite{hoeffding1994probability}, and the
bootstrap~\cite{efron1992bootstrap} to compute confidence intervals for both
\AlgBR and \AlgCR. For \AlgBR, we also consider the Clopper-Pearson
interval~\cite{clopper1934use}. For all settings, we use $\mathrm{Beta}(0.01, 1)$
data with a target recall of 90\%. As shown in Figure~\ref{fig:ci}, the normal
approximation matches or outperforms other methods within error margins. In
particular, Hoeffding's inequality does not use any property of the data in its
confidence interval (e.g., the variance) so it returns vacuous bounds. Since
Clopper-Pearson only applies to uniform sampling, we use the normal
approximation throughout to standardize confidence interval computation.

}

\begin{table}[t!]
\centering
\caption{\colora{Cost of \sn's query processing, executing the proxy model,
executing the oracle predicate, and exhaustive labeling. The oracle are human
labelers for all datasets except \texttt{night}, for which the oracle is an
expensive DNN (Table~\ref{table:datasets}). As shown, \sn's
query processing is orders of magnitude cheaper than the other parts of the
computation.}}
\vspace{-0.5em}
\label{table:costs}
\small
\begin{tabular}{l|llll|l}
   & \multicolumn{4}{c|}{\sn} & Exhaustive \\
  \specialcell{Dataset} & \specialcell{Sampling\\(AWS)} &
      \specialcell{Proxy\\(AWS)} & \specialcell{Oracle} & \specialcell{Total} &
      \specialcell{Oracle} \\
  \hline
  \texttt{night} & \$$1.7\times10^{-4}$ & \$0.02 & \$2.5 & \$2.52  & \$243 \\
  ImageNet       & \$$7.7\times10^{-5}$ & \$0.01 & \$80  & \$80.01 & \$4,000 \\
  OntoNotes      & \$$7.7\times10^{-5}$ & \$0.02 & \$80  & \$80.02 & \$893 \\
  TACRED         & \$$7.7\times10^{-5}$ & \$0.07 & \$80  & \$80.07 & \$1810
\end{tabular}
\vspace{-0.5em}
\end{table}

\subsection{Cost Analysis}
\colora{
We analyze the costs of query processing and executing the proxy/oracle methods.
For oracle predicates that are human labels, we approximate the cost by using
Scale API's \cite{scale} public costs, at \$0.08 per example. We approximate the cost of
computation by taking the cost per hour (\$3.06) of the Amazon Web Services
\texttt{p3.2xlarge} instance. The \texttt{p3.2xlarge} instance contains a single
V100 GPU and is commonly used for deep learning.

We show the breakdown of costs in Table~\ref{table:costs}. As shown, our
algorithms are significantly cheaper than exhaustive labeling. Furthermore, the
oracle predicate dominates the proxy models. Finally, \sn query processing costs
are negligible compared to the cost of both the proxy and oracle methods.

}

\section{Related Work}
\label{sec:related-work}

\minihead{Approximate query processing (AQP)}
AQP aims to return approximate answers to queries
for reduced computational complexity~\cite{hellerstein1997online,
agarwal2013blinkdb}. Most AQP systems focus on computing aggregates, such as
\texttt{SUM}~\cite{chakrabarti2001approximate, poosala1999fast},
\texttt{DISTINCT COUNT}~\cite{charikar2000towards, gibbons2001distinct,
haas1995sampling}, and quantiles~\cite{agarwal2013mergeable,
cormode2005improved,gan2018moment}.
Namely, these systems do not aim to answer selection queries. A
smaller body of work has studied approximate selection queries \cite{lazaridis2004approxselect}
with guarantees on precision and recall,
though to the best of our knowledge they do not provide probabilistic
guarantees or impose hard limits on usage of the predicate oracle,
and assume stronger semantics for the proxy.

\minihead{Optimizing relational predicates}
Researchers have proposed numerous methods of reducing the cost of relational
queries that contain expensive predicates~\cite{hellerstein1993predicate,
chaudhuri1999optimization, hellerstein1998optimization}. To the best of our
knowledge, this line of work does not consider approximate selection semantics.
In this work, we only consider a single proxy model and a single oracle model,
but existing optimization techniques may be useful if multiple oracle models must be applied.

\minihead{Information retrieval (IR), top-k queries}
IR and top-k queries typically aim to rank or select a
limited number of data points. Researchers have developed
exact~\cite{chang2002minimal, turtle1995query, broder2003efficient} and
approximate algorithms~\cite{anh2001vector, cambazoglu2010early} for these
queries. Other algorithms use proxy models for such
queries~\cite{wang2011cascade, gallagher2019joint}. To the best of our
knowledge, these methods and systems do not aim to do exhaustive selection.
Furthermore, we introduce notions of statistical guarantees on failure.

\minihead{Proxy models}
Approximate and proxy models have a long history in the machine learning
literature, e.g., cascades have been studied in the context of reducing
computational costs of classifiers~\cite{viola2001rapid, angelova2015real,
yang2016exploit}. However, these methods aim to maximize a single metric, such
as classification accuracy.

Contemporary visual analytics systems use a specific form of proxy model in the
form of specialized neural networks~\cite{kang2017noscope, kang2019blazeit,
lu2018accelerating, canel2019scaling, anderson2018predicate, hsieh2018focus}.
These specialized neural networks are used to accelerate queries, largely in the
form of binary detection~\cite{kang2017noscope, lu2018accelerating,
canel2019scaling, anderson2018predicate, hsieh2018focus}. We make use of these
models in our algorithms, but the choice and training of the proxy models
is orthogonal to our work.
Other systems use proxy models to accelerate other query types, such as selection
with \texttt{LIMIT} constraints or aggregation queries~\cite{kang2019blazeit}.

\section{Discussion and Future Work}
\label{sec:discussion}

While our novel algorithms for approximate selection queries with statistical
guarantees show promise, we highlight exciting areas of future work.

First, we have analyzed our algorithms in the asymptotic regime, in which the
number of samples goes to infinity. We believe finite-sample complexity bounds
will be a fruitful area of future research.

Second, we believe that information-theoretic lower bounds on sample complexity
are a fruitful area of future research. If these lower bounds on sample
complexity match the upper bounds from the algorithmic analysis, then these
algorithms are optimal up to constant factors. As such, we believe these bounds
will be helpful in informing future research.

Third, many scenarios naturally can have multiple proxy models. Our algorithms
have been developed for single proxy models and show the promise of
statistically improved algorithms for approximate selection with guarantees.
Furthermore, we believe these algorithms can improve statistical rates relative
to single proxy models in certain scenarios.

\section{Conclusion}
\label{sec:conclusion}

In this work, we develop novel, sample-efficient algorithms to execute
approximate selection queries \emph{with guarantees}. We define query
semantics for precision-target and recall-target queries with guarantees on
failure probabilities. We implement and evaluate our algorithms, showing that
they outperform existing baselines in prior work in all settings we
evaluated. These results indicate the promise of probabilistic algorithms to
answer selection queries with statistical guarantees. Supporting multiple
proxies and even more sample efficient algorithms are avenues for future
research.


\vspace{1em}

\subsection*{Acknowledgments}
{
\small
We thank Sahaana Suri, Kexin Rong, and members of the Stanford Infolab for their
feedback on early drafts. We further thank Tadashi Fukami, Trevor Hebert, and
Isaac Westlund for their helpful discussions. The hummingbird data was collected
by Kaoru Tsuji, Trevor Hebert, and Tadashi Fukami, funded by a Kyoto University
Foundation grant and an NSF Dimensions of Biodiversity grant (DEB-1737758). This
research was supported in part by affiliate members and other supporters of the
Stanford DAWN project---Ant Financial, Facebook, Google, Infosys, NEC, and
VMware---as well as Toyota Research Institute, Northrop Grumman, Amazon Web
Services, Cisco, and the NSF under CAREER grant CNS-1651570. Any opinions,
findings, and conclusions or recommendations expressed in this material are
those of the authors and do not necessarily reflect the views of the NSF. Toyota
Research Institute ("TRI") provided funds to assist the authors with their
research but this article solely reflects the opinions and conclusions of its
authors and not TRI or any other Toyota entity.

}

\vspace{1em}
\section{Additional Proofs}

\subsection{Theorem \ref{thm:opt_imp_sample}}
\label{sec:thm_imp_proof}

\begin{proof}
We decompose the variance conditioned on $a(x)$:
\begin{align*}
V \coloneqq &\var_{x \sim w}[f(x) u(x)/w(x)] \\
 = &\ex_{x \sim w}\left[\var_{x \sim w}[f(x)\frac{u(x)}{w(x)}|a(x)]\right] + \\
 & \var_{x \sim w}\left[\ex_{x \sim w}[f(x)\frac{u(x)}{w(x)}|a(x)]\right].
\end{align*}
Since $u(x),w(x)$ are known given $a(x)$ but $f(x)$ is not,
\begin{align*}
V &= \ex_{x \sim w} \left[a(x)(1-a(x))\frac{u(x)^2}{w(x)^2}\right] +
	\var_{x \sim w} \left[a(x)\frac{u(x)}{w(x)}\right] \\
  &= \ex_{x \sim w} \left[a(x) \frac{u(x)^2}{w(x)^2}\right] -
  	\ex_{x \sim w}\left[a(x)\frac{u(x)}{w(x)}\right]^2 \\
  &= \sum_{x} \left[a(x) \frac{u(x)^2}{w(x)}\right] - \ex_{x \sim u}\left[a(x)\right]^2
\end{align*}
In order to solve for the $w(x)$ minimizing $V$, we introduce the Lagrangian
dual for the constraint $\sum_x w(x) = 1$ and then take partial derivatives
of
\begin{equation*}
L \coloneqq \sum_{x} \left[a(x) \frac{u(x)^2}{w(x)}\right] - \ex_{x \sim u}\left[a(x)\right]^2 - \lambda\left(\sum_x w(x) - 1\right)
\end{equation*}
w.r.t. $w(x)$ to find that $-w(x)^{-2}a(x)u(x)^2 = \lambda$, so
$w(x) = C \sqrt{a(x)}u(x)$ for a normalizing constant $C$.
\end{proof}

\subsection{Variance comparisons}
\label{sec:varcompare}
Let
$
V_1 \coloneqq \sum_{x} \left[a(x) \frac{u(x)^2}{w(x)}\right] 
$ so that $V = V_1 - E_{u}[a(x)]^2$.
In this derivation we assume a uniform distribution $u(x)$. For uniform $w(x)$,
we have that
\begin{align*}
V_1^{(u)} = \sum_{x} a(x)u(x) = \ex_{x\sim u}[a(x)]
\end{align*}
For $w(x) \propto a(x)$:
\begin{align*}
  V_1^{(p)} &= \frac{1}{n^2}\sum_{\{x:a(x)>0\}} a(x)\cdot \frac{\sum_{x'}a(x')}{a(x)}\\
     &= \Pr(a(x)>0) \ex_{x\sim u}[a(x)]
\end{align*}
For $w(x) \propto \sqrt{a(x)}$:
\begin{align*}
V_1^{(s)} = \frac{1}{n^2}\sum_{x} a(x)\cdot\frac{\sum_{x'} \sqrt{a(x')}}{\sqrt{a(x)}} =
	\ex_{x\sim u}[\sqrt{a(x)}]^2
\end{align*}
We now show that these variances satisfy $V_1^{(s)} \leq  V_1^{(p)} \leq V_1^{(u)} $.

First, note that $\Pr(a(x)>0) \leq 1$ implies $V_1^{(p)} \leq V_1^{(u)}$.

Using H\"older's inequality, we have that
\[\ex_{x\sim u}[\sqrt{a(x)}1_{a(x)>0}] \leq \ex_{x\sim u}[a(x)]^{1/2}\ex_{x\sim u}[1_{a(x)>0}]^{1/2}.\]
Squaring both sides yields
\[\ex_{x\sim u}[\sqrt{a(x)}]^2 \leq \ex_{x\sim u}[a(x)]\Pr(a(x)>0).\]

Finally, note that the gap between the optimal and uniform weights has a simple form
\[
V_1^{(u)} - V_1^{(s)} = \var_{x\sim u}[\sqrt{a(x)}].
\]

\vspace{1em}

\balance

\bibliographystyle{abbrv}
\Urlmuskip=0mu plus 1mu

\bibliography{paper}

\clearpage

\iftoggle{arxiv}{
}{}
\begin{appendix}
\section{Queries with Both a Precision and Recall Target}
\label{sec:jt}

Certain applications may require both precision and recall targets
simultaneously. We refer to queries that impose both
targets as JT queries. Unlike with PT and RT queries,
achieving these targets may not be possible given a maximum
oracle label budget. In our experience, industry and research
applications usually prefer queries that can operate under a budget
so JT queries are less practical.
Nonetheless, we discuss syntax and semantics of such queries, describe
algorithms that provide guarantees, and evaluate these algorithms.

\begin{figure}
\begin{lstlisting}[frame=single]
SELECT * FROM table_name
  WHERE filter_predicate
  USING proxy_estimates
  RECALL TARGET tr
  PRECISION TARGET tp
  WITH PROBABILITY p
\end{lstlisting}
  \caption{Syntax for specifying approximate selection queries with both a
  precision and recall target. Users provide both a precision and recall target
  in addition to a probability of success. Such queries do not have an oracle
  budget, as the oracle may be queried an unbounded number of times.}
  \label{fig:syntax-both}
\end{figure}

\minihead{Syntax and semantics}
We define the syntax for queries with both a recall and precision target in
Figure~\ref{fig:syntax-both}. Such queries do not specify an oracle budget and
instead specify both a precision and recall target. All other syntax elements
are a direct extension of the syntax for PT and RT queries in Figure~\ref{fig:syntax}.

The semantics of such queries remains similar, except the recall and precision
targets must be jointly satisfied with a failure probability.

\minihead{Algorithms}
We describe how the recall target algorithms can be used as a subroutine for the
joint target (JT) setting. We note a modified version of the precision target
algorithm can be used as a subroutine instead, but defer the analysis and
evaluation to future work.

The joint algorithm proceeds in three stages. 
\begin{enumerate}
  \item Optimistically allocate a budget $B$.
  \item Run an RT algorithm (\AlgCR) with budget $B$ to achieve the recall
  target $\tm_r$.
  \item Exhaustively filter out false positives from stage two using oracle
  queries.
\end{enumerate}

This algorithm achieves the recall target because the RT algorithm 
will return a set of records with sufficient positive matches with high probability. 
Then, since we exhaustively query the oracle on the results from stage two in the third stage,
we can prune the results to achieve a precision target while maintaining the recall.
Note that the third stage is necessary because even with a large budget for estimating
a proxy score threshold $\tm_r$, there may not exist any threshold that would allow
us to directly identify a set of records with high enough precision and recall.
The proxy may simply be too weak to identify records with the required precision
and recall targets, so any JT algorithm will have to supplement sets of records
identified by the proxy with additional processing.

\minihead{Evaluation}
We evaluate the joint algorithm when using uniform sampling and our improved
importance sampling recall target algorithms as subroutines.
For comparison, we use the same budget as our evaluations in Section~\ref{sec:eval_fu}
to run stage 2 of the JT method.

\begin{figure}[t!]
  \includegraphics[width=\columnwidth]{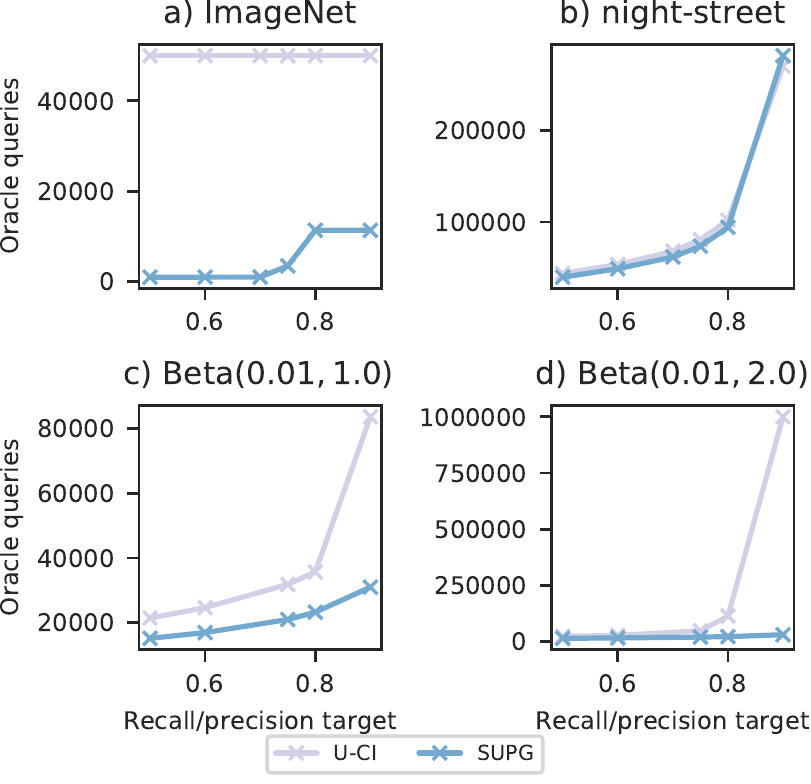}
  \caption{\sn and uniform sampling on joint target queries. As shown, \sn
  generally outperforms uniform sampling.}
  \label{fig:fu-jt}
\end{figure}

We target precision/recall targets of 0.5, 0.6, 0.7, 0.75, 0.8, and 0.9 and
executed the joint algorithms. We show the recall/precision target vs the number
of oracle queries in Figure~\ref{fig:fu-jt}. As shown, \sn generally outperforms
uniform sampling.

\end{appendix}


\end{document}